\documentclass[twocolumn,showkeys,showpacs,preprintnumbers,prd,superscriptaddress,nofootinbib]{revtex4-1}
\usepackage{bold-extra}
\usepackage{graphicx}
\usepackage{epsf}
\usepackage{bm}
\usepackage{amsmath}
\usepackage{amsfonts}
\usepackage{booktabs}
\usepackage{amssymb}
\usepackage{epstopdf}
\usepackage{natbib}
\usepackage{color}
\usepackage[dvipsnames,table]{xcolor}
\usepackage{physics}
\usepackage[colorlinks=true,linkcolor=Blue,citecolor=blue,urlcolor=magenta]{hyperref}
\usepackage{verbatim}
\usepackage{lipsum}
\usepackage{float}
\usepackage{multirow}
\usepackage{soul}

\usepackage[capitalize]{cleveref}
\providecommand{\U}[1]{\protect\rule{.1in}{.1in}}

\newcommand{\be}{\begin{equation}}
	\newcommand{\ee}{\end{equation}}
\newcommand{\bea}{\begin{eqnarray}}
	\newcommand{\eea}{\end{eqnarray}}

\usepackage{tikz}

\begin{document}

	\title{Dark Energy Crosses the Line: Quantifying and Testing the Evidence for Phantom Crossing}
	
	\author{Emre \"{O}z\"{u}lker}
	\email{e.ozulker@sheffield.ac.uk}
	\affiliation{School of Mathematical and Physical Sciences, University of Sheffield, Hounsfield Road, Sheffield S3 7RH, United Kingdom} 
	
	\author{Eleonora Di Valentino}
	\email{e.divalentino@sheffield.ac.uk}
	\affiliation{School of Mathematical and Physical Sciences, University of Sheffield, Hounsfield Road, Sheffield S3 7RH, United Kingdom} 
	
	\author{William Giar\`e}
	\email{giare@hawaii.edu}
	\affiliation{Department of Physics and Astronomy, University of Hawai‘i, Honolulu, HI 96822, USA}

	\begin{abstract}
		Combinations of the most recent CMB, BAO, and SNeIa datasets, when analyzed using the CPL parametrization, $w(a) = w_0 + (1 - a) w_a$, exclude $\Lambda$CDM at $\gtrsim\!3\sigma$ in favor of a dark energy equation of state (EoS) parameter that crosses the phantom divide. We confirm this behavior and show that it persists when DESI BAO data are replaced by SH0ES $H_0$ measurements, despite the known tension between these probes in the presence of CMB data. In both cases, the constraints favor a transition from an early-time phantom-like phase to a late-time quintessence-like phase, with the crossing occurring at different redshifts depending on the dataset combination. The probability that a phantom divide line (PDL) crossing does not occur within the expansion history is excluded at significance levels ranging \mbox{between $3.1\sigma$\,--\,$5.2\sigma$.} To investigate whether the apparent PDL crossing is a genuine feature preferred by the data or an artifact of the linear form of the CPL parametrization, we isolate the PDL crossing feature by introducing two modified versions of CPL that explicitly forbid it: CPL${}_{>a_\mathrm{c}}$ and CPL${}_{<a_\mathrm{c}}$. These models are physically motivated in that they phenomenologically capture the behavior of thawing and freezing scalar field scenarios. While previous studies have primarily considered thawing quintessence as a non-crossing alternative, we find that a freezing phantom-like model is the only one capable of performing comparably to CPL—and only in a few cases. Nevertheless, across all dataset combinations considered, the standard CPL model consistently provides the best fit, strongly indicating that the data genuinely favor a PDL crossing.
	\end{abstract}
	\maketitle

	\section{Introduction}
	The standard cosmological paradigm, the $\Lambda$CDM model, is marked by two components of elusive physical origin, namely, the cosmological constant ($\Lambda$) and cold dark matter (CDM). The initial remarkable accuracy of the $\Lambda$CDM model in explaining the cosmological observations had provided strong motivation to look for theoretical frameworks underlying these components. However, the persistence and growing statistical significance of various observational tensions—emerging with the increasing precision of data over the past decade—suggests the need for a new cosmological framework that can more accurately describe the observed Universe~\cite{Abdalla:2022yfr,Perivolaropoulos:2021jda,DiValentino:2022fjm,CosmoVerse:2025txj}. 
	
	The most prominent of the tensions is in the value of the Hubble constant, $H_0$~\cite{Verde:2019ivm,DiValentino:2020zio,DiValentino:2021izs,Perivolaropoulos:2021jda,Schoneberg:2021qvd,Shah:2021onj,Abdalla:2022yfr,DiValentino:2022fjm,Kamionkowski:2022pkx,Giare:2023xoc,Hu:2023jqc,Verde:2023lmm,DiValentino:2024yew,CosmoVerse:2025txj}. Distance ladder measurements of $H_0$ relying on the calibration of Type Ia supernovae (SNeIa) by Cepheid variables are in $\gtrsim\!5\sigma$ tension with the $H_0$ values~\cite{Riess:2021jrx,Murakami:2023xuy,Breuval:2024lsv} based on cosmic microwave background (CMB) observations that are inferred by assuming $\Lambda$CDM cosmology~\cite{Planck:2018vyg,ACT:2025fju,SPT:2023jql}. In particular, the SH0ES team reports a distance ladder measurement of $H_0=73.04\pm1.04$~km s${}^{-1}$ Mpc${}^{-1}$~\cite{Riess:2021jrx} whereas the CMB temperature, polarization, and lensing data of \textit{Planck} collaboration yields $H_0=67.36\pm0.54$~km s${}^{-1}$ Mpc${}^{-1}$ when the $\Lambda$CDM model is assumed~\cite{Planck:2018vyg}. While there are other numerous late-time inferences of $H_0$ 
	that are $\Lambda$CDM independent~\cite{Freedman:2020dne,Birrer:2020tax,Anderson:2023aga,Scolnic:2023mrv,Jones:2022mvo,Anand:2021sum,Freedman:2021ahq,Uddin:2023iob,Huang:2023frr,Li:2024yoe,Pesce:2020xfe,Kourkchi:2020iyz,Schombert:2020pxm,Blakeslee:2021rqi,deJaeger:2022lit,Freedman:2024eph,Riess:2024vfa,Vogl:2024bum,Scolnic:2024hbh,Said:2024pwm,Boubel:2024cqw,Scolnic:2024oth,Li:2025ife,Jensen:2025aai} some of which yield a reduced statistical significance of the tension, concordance on the value of $H_0$ cannot be achieved by assuming unknown systematic errors in one or few late-time probes. Even for different teams, objects, and calibrators, the tension still varies between $4$--$6\sigma$~\cite{Riess:2019qba,DiValentino:2020vnx,DiValentino:2022fjm}. For discussions of possible systematic effects on the data, see also~\cite{Dominguez:2019jqc,Boruah:2020fhl,Mortsell:2021nzg,Mortsell:2021tcx,Riess:2021jrx,Sharon:2023ioz,Murakami:2023xuy,Riess:2023bfx,Bhardwaj:2023mau,Brout:2023wol,Dwomoh:2023bro,Uddin:2023iob,Riess:2024ohe,Freedman:2024eph,Riess:2024vfa}.
	
	Recently, a new challenge surfaced in the form of a $\sim\!3.1\sigma$ preference for an evolving dark energy (DE) density over the cosmological constant when \textit{Planck} CMB measurements are combined with baryon acoustic oscillations (BAO) measurements of the DESI collaboration~\cite{DESI:2024mwx, DESI:2025zgx}. 
	Such preference is strengthened up to $\sim\!4.2\sigma$ when the uncalibrated Dark Energy Survey (DES) Y5 SNeIa~\cite{DES:2024jxu} 
	are also added to the data set~\cite{DESI:2025zgx}. This preference was found when the DE dynamics were allowed to manifest within the Chevallier-Polarski-Linder (CPL) parameterization~\cite{Chevallier:2000qy, Linder:2002et}, which replaces the cosmological constant with a phenomenological DE component whose equation of state (EoS) parameter is a linear function of time. It is worth noting that the combination of the DESI BAO with uncalibrated SNeIa data sets without any information from CMB also has a preference for the CPL parameterization~\cite{Giare:2025pzu,DESI:2025zgx}. However, for the Pantheon\textsuperscript{+} SNeIa compilation, this is low enough to be consistent with a cosmological constant within $2\sigma$.\footnote{Also, when CMB data different from \textit{Planck} are used, the statistical significance of the above findings is reduced~\cite{Giare:2024ocw} and even removed for certain data set combinations including simultaneously SDSS BAO data and the Pantheon\textsuperscript{+} SNeIa sample~\cite{Giare:2025pzu}.} Thus, evidence for dynamical DE is present in all three cases of combining DESI BAO with only CMB, only SNeIa, or both. Finally, this result remains robust under alternative choices of DDE parameterizations and dataset combinations~\cite{DESI:2024mwx,Cortes:2024lgw,Shlivko:2024llw,Luongo:2024fww,Yin:2024hba,Gialamas:2024lyw,Dinda:2024kjf,Najafi:2024qzm,Wang:2024dka,Ye:2024ywg,Tada:2024znt,Carloni:2024zpl,Chan-GyungPark:2024mlx,DESI:2024kob,Bhattacharya:2024hep,Ramadan:2024kmn,Notari:2024rti,Orchard:2024bve,Hernandez-Almada:2024ost,Pourojaghi:2024tmw,Giare:2024gpk,Reboucas:2024smm,Giare:2024ocw,Chan-GyungPark:2024brx,Menci:2024hop,Li:2024qus,Li:2024hrv,Notari:2024zmi,Gao:2024ily,Fikri:2024klc,Jiang:2024xnu,Zheng:2024qzi,Gomez-Valent:2024ejh,RoyChoudhury:2024wri,Lewis:2024cqj,Wolf:2025jlc,Shajib:2025tpd,Giare:2025pzu,Chaussidon:2025npr,Kessler:2025kju,Pang:2025lvh,RoyChoudhury:2025dhe,Scherer:2025esj,Specogna:2025guo,Cheng:2025lod,Nesseris:2025lke,Akrami:2025zlb,Cheng:2025hug,Gonzalez-Fuentes:2025lei,Ye:2024zpk,Sabogal:2025jbo,Gomez-Valent:2025mfl,Lu:2025gki}.\footnote{At the same time, the interpretation of the DESI results is still debated. Questions have been raised about the statistical interpretation of overlapping SNeIa compilations~\cite{Cortes:2025joz}, the consistency of combined datasets~\cite{Wang:2025bkk}, and possible inconsistencies between BAO and SNeIa distances~\cite{Afroz:2025iwo}. The influence of potential outliers in the DESI DR1 BAO sample has also been examined~\cite{Sapone:2024ltl}.}

	The combination of all three suggests not only a dynamical DE component but also one whose EoS parameter has the unconventional feature of crossing the phantom divide line (PDL).

	For a perfect fluid with positive energy density, a phantom-like EoS implies $\rho+p<0$. For a canonical, minimally coupled scalar field $\phi$, however, $\rho_\phi+p_\phi=\dot{\phi}^2\geq0$, which prevents both a phantom phase and a PDL crossing. Reversing the sign of the kinetic term permits $w<-1$ but introduces a ghost degree of freedom. A crossing therefore requires dynamics beyond canonical quintessence, for instance additional degrees of freedom or nonminimal coupling~\cite{Cai:2009zp,Ye:2024ywg}.\footnote{Concrete realizations involving noncanonical scalar dynamics or interactions between DE and dark matter have also been investigated, see, e.g.,~\cite{Giare:2024gpk,Chen:2025ywv,Guedezounme:2025wav,Li:2026xaz}.} The EoS inferred from the expansion history may then describe an effective DE component rather than the intrinsic EoS of a single field.\footnote{Consequently, a preference for a PDL crossing does not identify its physical origin, and the stability of any proposed realization must be assessed within that theory.}

	The evidence for the PDL crossing feature in the presence of the DESI BAO data appears to be robust across different approaches. It is observed not only within the CPL parametrization, but also in analyses employing alternative functional forms of the dark energy EoS parameter or the DE density~\cite{Wolf:2024stt,Scherer:2025esj,Wolf:2025jed,Nesseris:2025lke,Gomez-Valent:2025mfl}, as well as in model-independent reconstructions of the DE density and/or equation-of-state parameter~\cite{DESI:2024aqx,Ye:2024ywg,Gonzalez-Fuentes:2025lei}.
	\footnote{Nevertheless, the interpretation of the inferred crossing remains under discussion. Monte Carlo tests have examined whether a CPL preference for crossing can arise from statistical fluctuations when the underlying model does not cross the PDL~\cite{Keeley:2025rlg}. Cosmographic analyses have also found evolving DE without a crossing~\cite{Roy:2025cxk}.}

	Here, we focus on the PDL crossing feature found within the CPL parametrization in relation to the $H_0$ tension, and check the robustness of the evidence by analyzing the data with modified versions of CPL that do not allow a crossing of the PDL.
	
	The paper is structured as follows. In Sec.~\ref{sec:CPL}, we investigate the PDL crossing in the standard CPL parametrization. In Sec.~\ref{sec:ModifiedCPL}, we introduce two modified CPL models that do not allow a PDL crossing. In Sec.~\ref{sec:Observations}, we present the observational constraints on these models using various data combinations. Finally, we draw our conclusions in Sec.~\ref{sec:Conclusion}.

	\section{PDL Crossing within CPL parametrization} \label{sec:CPL}
	
	\begin{figure*}
		\centering
		\begin{tikzpicture}
			\node (img) {
				\includegraphics[width=0.5\textwidth]{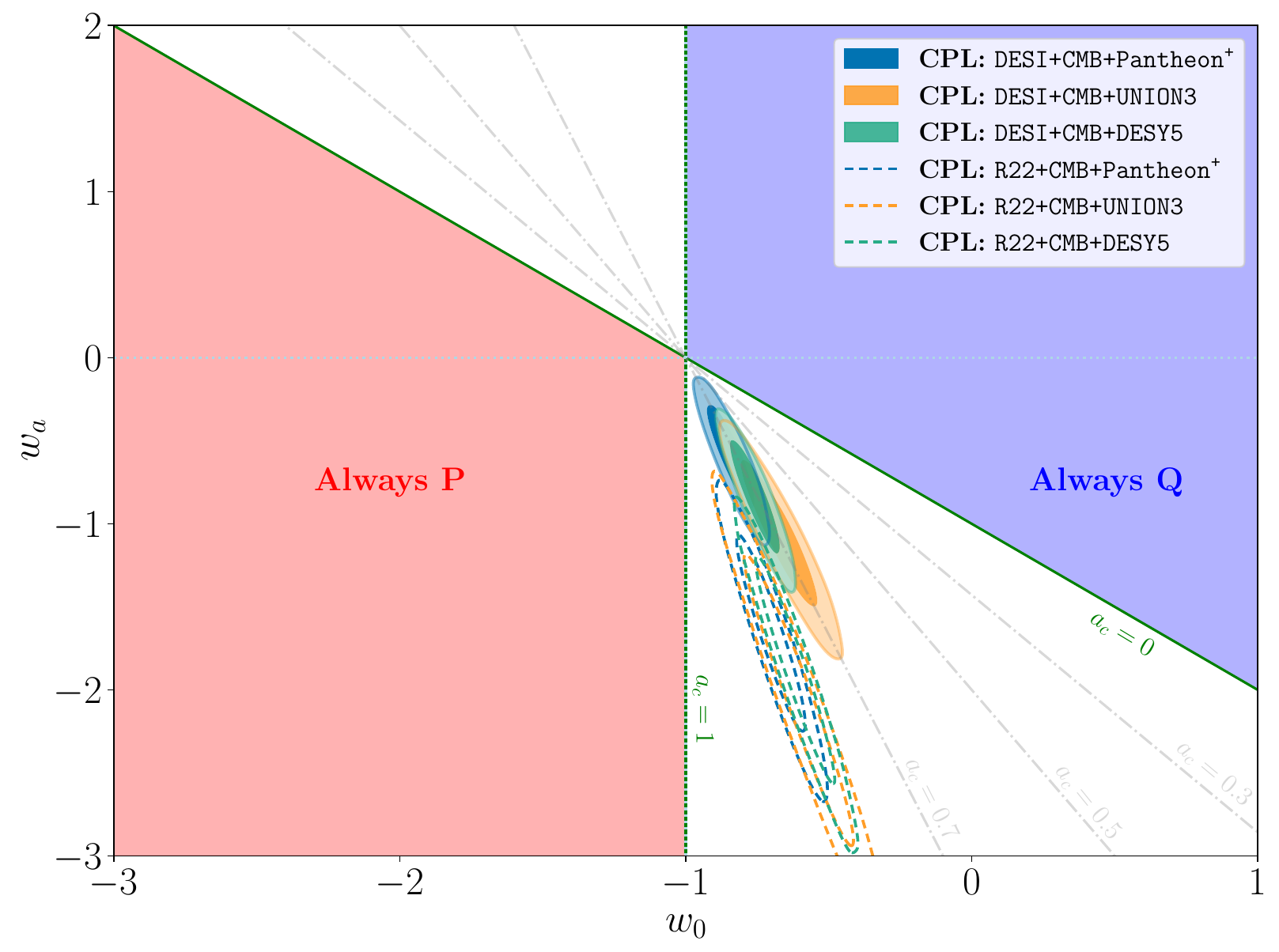}
				\raisebox{-0.01cm}{\includegraphics[width=0.489\textwidth]{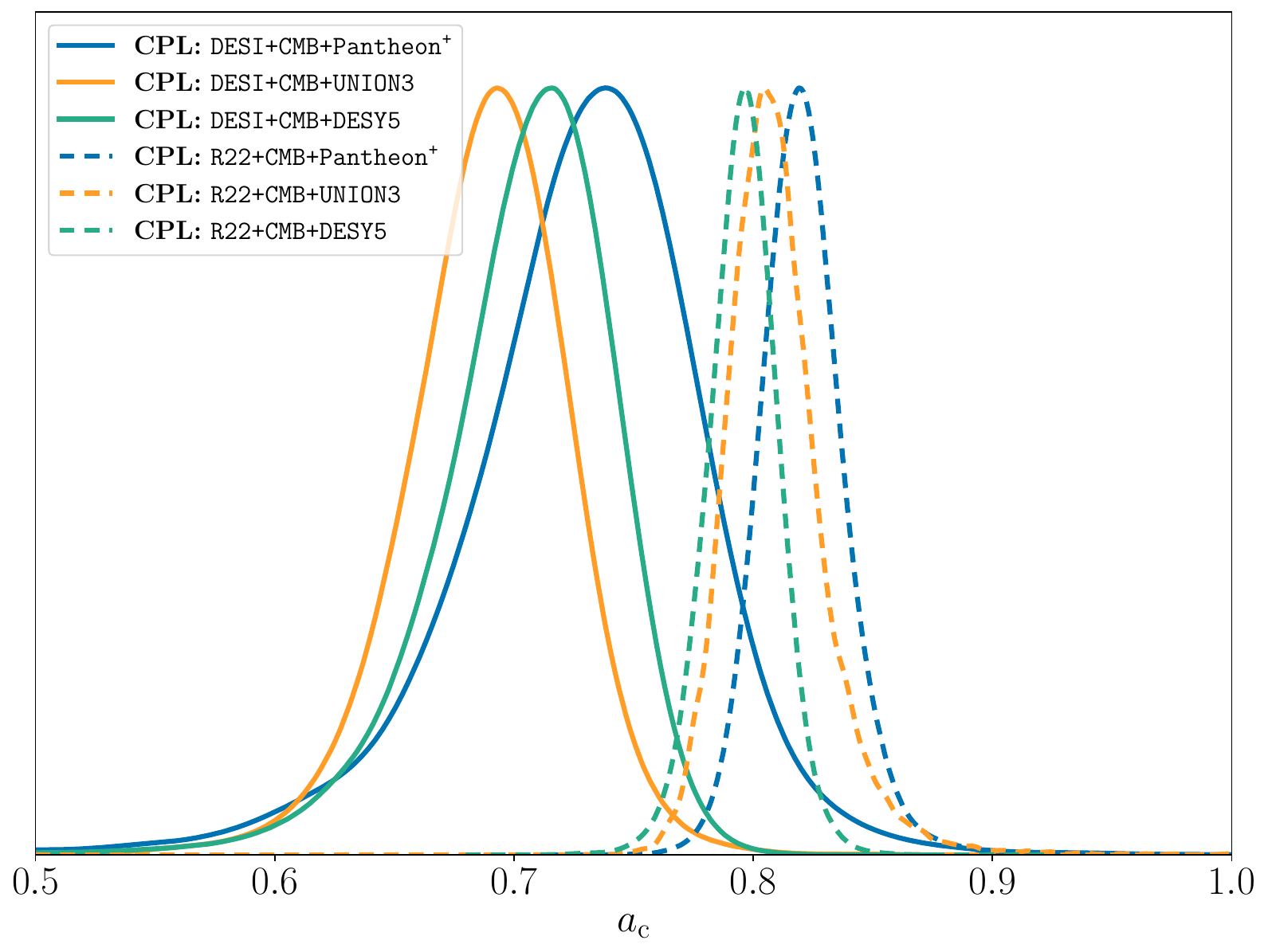}}
			};
			\begin{scope}[shift={(-5, 2.25)},scale=0.5,transform shape]
				\node[] at (-2.5ex,0) {PDL};
				\node[] at (0.7ex,2.5ex) {Q};
				\node[] at (7.3ex,-2.5ex) {P};
				\draw [-stealth](1.5ex,1.25ex) -- (6.5ex,-1.75ex);
				\draw (0,0) -- (8ex,0);
			\end{scope}
			\begin{scope}[shift={(-1.75, -1.75)},scale=0.5,transform shape]
				\node[] at (-2.5ex,0) {PDL};
				\node[] at (7ex,2.8ex) {Q};
				\node[] at (0.9ex,-2.2ex) {P};
				\draw [-stealth](1.5ex,-1.25ex) -- (6.5ex,1.75ex);
				\draw (0,0) -- (8ex,0);
			\end{scope}
		\end{tikzpicture}
		\caption{\textit{Left panel:} 2D contour plots based on a baseline combination of CMB+SNeIa data. Solid contours correspond to results obtained by adding DESI DR2 BAO data, while dashed empty contours represent results using \texttt{R22} instead of \texttt{DESI}. In this figure, regions without a PDL crossing are colored red if they always correspond to a phantom equation of state, or blue if they always correspond to quintessence. In the white regions where a PDL crossing occurs, the direction of the crossing is indicated with an arrow. In the top left, it corresponds to a PDL crossing from a quintessence-like EoS parameter to a phantom-like one as the Universe expands (Q$\to$P), while in the bottom right, it is the opposite (P$\to$Q). \textit{Right panel:} 1D marginalized posteriors of $a_\mathrm{c}$ for the same dataset combinations as in the left panel. }
		\label{fig:DESI-like}
	\end{figure*}
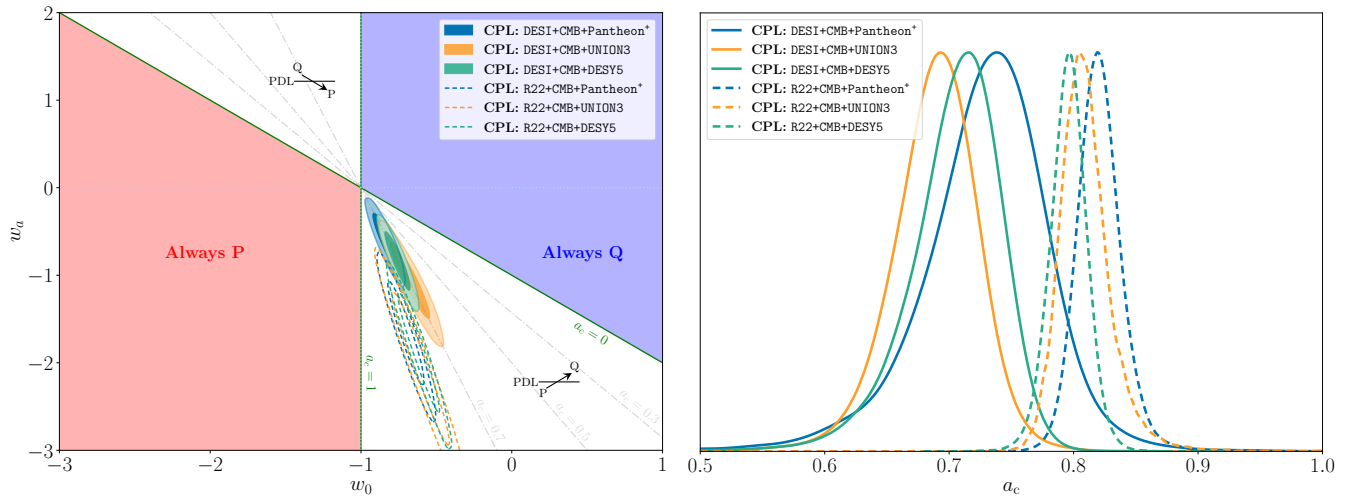

	Gravitational aspects of a perfect fluid DE are characterized by its energy density, $\rho$, and pressure, $p$. In a spatially uniform Universe described by the Robertson–Walker (RW) metric with scale factor $a$, the local conservation of energy-momentum implies the continuity equation relating these functions as follows:
	\begin{equation}
		\dv{\rho}{a}=-3\frac{\rho+p}{a}.
		\label{eq:cont}
	\end{equation}
	One can define the EoS parameter, $w \equiv p/\rho$, and~\cref{eq:cont} can be integrated to write
	\begin{equation}
		\abs{\rho(a_2)} = \abs{\rho(a_1)} \exp\left[-3\int_{a_1}^{a_2} \dd{a} \frac{1+w(a)}{a} \right].
	\end{equation}
	The above equation shows how the dynamics of the DE fluid are encoded in the EoS parameter. Since the cosmological constant is effectively a perfect fluid with $w = -1$, any EoS parameter deviating from this constant function would also capture deviations from the $\Lambda$CDM model.
	
	The CPL parametrization~\cite{Chevallier:2000qy, Linder:2002et} models the EoS parameter as a linear function of the scale factor,
	\be
	w(a) = w_0 + (1 - a) w_a,
	\label{eq:cpl}
	\ee
	where $w_0$ denotes the present-day value of the EoS parameter in our convention of $a_0 = 1$, with the subscript “0” denoting the present-day value of any quantity henceforth. This simple parametrization can imitate or describe major cosmological observables of a diverse set of DE models with an accuracy at the level of $10^{-3}$ or better~\cite{dePutter:2008wt}. Moreover, it is useful in capturing deviations from the $\Lambda$CDM model, since $\Lambda$CDM is nested within the CPL model.\footnote{However, note that it cannot capture some nontrivial deviations, such as an effective DE density that attains negative values in the past, which would require a singular EoS parameter~\cite{Ozulker:2022slu}.}
	
	An interesting feature of the CPL parametrization is that it can easily realize a scenario in which the DE EoS parameter crosses the PDL ($w = -1$) in either direction: from a quintessence-like EoS to a phantom-like one, or vice versa. The scale factor at which the PDL crossing occurs, which we denote as $a_{\rm c}$, must satisfy
	\be
	w(a_{\rm c}) = -1.
	\label{eq:ac_sat}
	\ee
	In fact, there is always a solution
	\be
	a_{\rm c} = 1 + \frac{1 + w_0}{w_a},
	\label{eq:ac_eq}
	\ee
	provided that $w_a \neq 0$.\footnote{In the special case of a cosmological constant with $w_a = 0$ and $w_0 = -1$, there are infinitely many solutions, but of course, they do not correspond to a PDL crossing.} However, the CPL parametrization incorporates the PDL crossing feature meaningfully only when $a_{\rm c} \in [0,1]$. Otherwise, one either obtains a nonphysical value with $a_{\rm c} < 0$, or a predicted PDL crossing in the future with $a_{\rm c} > 1$. 
	Considering that \cref{eq:cpl} does not arise from fundamental physics but is instead a first-order approximation to a more realistic DE EoS parameter, the CPL parametrization should be treated as a fitting model rather than a predictive one. Hence, if observational data favor $a_{\rm c} > 1$, this indicates evidence \textit{against} a PDL crossing in the DE EoS parameter. It is also important to note that even if observational data prefer $a_{\rm c} \in [0,1]$, this may simply be an artifact of the parametrization, as discussed further in the following sections.
	
	The $w_0$–$w_a$ plane can be divided into four regions based on the behavior of the EoS parameter with respect to the PDL. In the left panel of~\cref{fig:DESI-like}, we show these four regions, using the label “Q” to indicate a quintessence-like EoS parameter satisfying $w > -1$, and the label “P” to indicate a phantom-like one satisfying $w < -1$. The shaded blue region is where $w(a) > -1$ holds for all $a \in [0,1]$, and it is the most compelling from the point of view of a single scalar field. The shaded red region is where $w(a) < -1$ holds for all $a \in [0,1]$. The two white regions correspond to PDL crossings in opposite directions. The white region in the top left corresponds to a PDL crossing from a quintessence-like EoS parameter to a phantom-like one as the Universe expands; it is denoted by Q$\to$P. The final region, in the bottom right—of particular interest to us due to the observational constraints discussed below—corresponds to a PDL crossing from a phantom-like EoS parameter to a quintessence-like one as the Universe expands; this region is denoted by P$\to$Q.
	
	Notice that these four regions are separated by the green lines corresponding to $a_{\rm c} = 1$ and $a_{\rm c} = 0$. As can be seen from~\cref{eq:ac_eq}, a given value of $a_{\rm c}$ corresponds to a line in the $w_0$–$w_a$ plane whose slope is $1/(a_{\rm c} - 1)$. Therefore, a strong correlation between the parameters $w_0$ and $w_a$ would naturally lead to a precise determination of $a_{\rm c}$ as also seen from the gray dash-dotted lines in the left panel of~\cref{fig:DESI-like}, showing three example cases where $a_{\rm c}$ takes the values of 0.3, 0.5, and 0.7. Moreover, all lines of constant $a_{\rm c}$ intersect at the vertex point $(w_0 = -1, w_a = 0)$, which corresponds to the cosmological constant. This is also evident in the left panel of~\cref{fig:DESI-like}, where this point separates all four regions, analogous to a quadruple point on a phase diagram. The white regions—being the regions of PDL crossing—contain all the lines for $0 < a_{\rm c} < 1$, while the remaining lines, corresponding to $a_{\rm c} > 1$ or $a_{\rm c} < 0$, lie within the shaded red and blue regions. It is noteworthy that, across all dataset combinations, the direction of the correlation in the $w_0$–$w_a$ plane aligns with the cosmological constant along the PDL crossing lines, underscoring the robustness of this preference.

	\section{Modified CPL models without PDL crossing: CPL\boldmath{${}_{>a_\mathrm{c}}$} and CPL\boldmath{${}_{<a_\mathrm{c}}$} } \label{sec:ModifiedCPL}
	
	The standard CPL parametrization permits a smooth crossing of the PDL as a result of its linear form. While this allows a broad class of EoS behaviors to be modeled within a simple two-parameter framework, it also raises the question of whether the preference for a PDL crossing observed in data analyses truly reflects a physical feature requiring explanation, or whether it is merely a consequence of the functional form of the CPL parametrization.
	
	In particular, the linear nature of CPL imposes a fixed slope on $w(a)$ at all times. Therefore, even if the data merely prefer a varying EoS parameter within a narrow redshift window—without requiring an actual PDL crossing—the model may still accommodate (or even favor) such a crossing as an artifact. This motivates the construction of modified models that preserve the ability to describe dynamical DE around the crossing scale factor $a_{\rm c}$, while deliberately preventing the EoS parameter from crossing the PDL.
	
	We therefore define two physically motivated modifications of the CPL parametrization that are devoid of PDL crossings, by imposing $w(a) = -1$ where the usual CPL form would otherwise cross to the opposite side of the phantom divide. We refer to these models as CPL${}_{>a_\mathrm{c}}$ and CPL${}_{<a_\mathrm{c}}$, where the subscripts indicate the region of dynamical behavior. It is important to emphasize that these modifications do not introduce additional free parameters, as $a_\mathrm{c}$ is entirely determined by $w_0$ and $w_a$, as shown in~\cref{eq:ac_eq}. Moreover, they are completely equivalent to CPL when $a_\mathrm{c}\notin [0,1]$, in which case PDL crossing is absent in all three models, and they exhibit either quintessence or phantom behavior at all times throughout the expansion history.  More precisely, the two modifications we introduce are:
	\begin{itemize}
		\item \textbf{CPL}${}_{\bm{>a_\mathrm{c}}}$\textbf{:} The EoS parameter behaves as a cosmological constant, $w = -1$, in the early Universe. It transitions into the CPL form only once the scale factor reaches the would-be PDL crossing scale $a_{\rm c}$, and then follows the CPL evolution from that point onward. Hence, the EoS parameter is given by
		\begin{equation}
			w(a) = 
			\begin{cases}
				-1, & \text{for } a < a_\mathrm{c} \\
				w_0 + (1 - a) w_a, & \text{for } a \geq a_\mathrm{c}.
			\end{cases}
		\end{equation}
		
		\item \textbf{CPL}${}_{\bm{<a_\mathrm{c}}}$\textbf{:} The EoS parameter follows the CPL form initially, up to the scale factor $a_{\rm c}$ at which a PDL crossing would occur. From that point on, it is held fixed at $w = -1$, freezing the EoS parameter to mimic a cosmological constant at late times. Hence, the EoS parameter is given by
		\begin{equation}
			w(a) = 
			\begin{cases}
				w_0 + (1 - a) w_a, & \text{for } a < a_\mathrm{c} \\
				-1, & \text{for } a \geq a_\mathrm{c}.
			\end{cases}
		\end{equation}
	\end{itemize}
	
	These two models mimic the behavior of CPL either only before or only after the point where a crossing would occur in CPL ($a_\mathrm{c}$); during the remainder of the cosmic evolution, they are frozen and mimic a cosmological constant. Qualitatively, both of these modifications—each incorporating one dynamical and one static phase for the DE density—resemble well-known behaviors of single scalar field scenarios. 
	For instance, CPL${}_{>a_\mathrm{c}}$ describes a DE density that remains constant for $a < a_\mathrm{c}$ but begins evolving for $a > a_\mathrm{c}$. This is reminiscent of a slow-roll field~\cite{Linde:1981mu,Albrecht:1982wi} that exits its slow-rolling phase at $a \sim a_\mathrm{c}$, or more relevantly for DE, a thawing quintessence field~\cite{Caldwell:2005tm}. In fact, see Ref.~\cite{Notari:2024rti} for a concrete potential that can realize CPL${}_{>a_\mathrm{c}}$ to arbitrary precision, and Ref.~\cite{Wolf:2024eph} for a smooth and generic potential with very similar background phenomenology.
	Similarly, CPL${}_{<a_\mathrm{c}}$ describes an initially evolving DE density that freezes at $a = a_\mathrm{c}$, reminiscent of freezing quintessence models~\cite{Caldwell:2005tm}. Of course, these correspondences assume that the phenomenology of the scalar field in its dynamical phase is well approximated by a first-order expansion of its EoS parameter, i.e.,~\cref{eq:cpl}.
	Note that these correspondences concern the background evolution of the DE density and EoS parameter. Their interpretation in terms of canonical thawing or freezing quintessence applies to branches with $w>-1$. The phantom-like branches of the modified models require a different physical realization.
	
	These constructions serve to isolate whether the fit improvement seen in CPL analyses stems from the PDL crossing itself or from the ability of the model to accommodate local slope variations in $w(a)$. If either CPL${}_{>a_\mathrm{c}}$ or CPL${}_{<a_\mathrm{c}}$ yields comparable or better agreement with observations, it would suggest that the evidence for a PDL crossing in CPL may be an artifact—not because PDL crossing is essential for describing the observations, but because it arises as a byproduct of the features of CPL that are essential for describing the observations.

	\section{Evidence for a PDL crossing from observations} \label{sec:Observations}
	
	We confront the CPL model with observational data to investigate whether a PDL crossing is preferred. To this end, we adopt the three dataset combinations used by the DESI collaboration in Refs.~\cite{DESI:2024mwx,DESI:2025zgx}, which yield the strongest preference for dynamical DE in their analyses. These combinations include CMB and BAO data together with three different SNeIa compilations. In this context, the unanchored BAO measurements are calibrated by the stringent constraints on the sound horizon from CMB data, which can then be combined with uncalibrated SNeIa.
	In addition, we consider three alternative dataset combinations by replacing the DESI BAO data in the previous sets with distance ladder measurements of the Hubble constant $H_0$ from the SH0ES team. In this second case, the SNeIa data are calibrated using the distance ladder, which can be consistently combined with CMB data.
	Furthermore, we construct two modified versions of the CPL model in which PDL crossing is explicitly forbidden, and we confront these modified CPL models with the same six dataset combinations.

	\subsection{Methodology and Data}
	
	We compute the cosmological observables for all three models using a modified version of the Boltzmann code \texttt{CAMB}~\cite{Lewis:1999bs}, which we adapt to implement the two CPL variations that explicitly forbid a PDL crossing. These observables are then passed to the \texttt{Cobaya} framework~\cite{Torrado:2020dgo} to perform Bayesian parameter inference via Markov Chain Monte Carlo (MCMC) sampling using its \texttt{mcmc} sampler.\footnote{Documentation available at \url{https://cobaya.readthedocs.io/en/latest/sampler_mcmc.html}.} 
	For modeling DE perturbations, we employ the default parameterized post-Friedmann (PPF) prescription in \texttt{CAMB}~\cite{Hu:2007pj,Fang:2008sn}, which permits a regular treatment of the perturbations when $w(a)$ approaches or crosses $-1$. This prescription describes perturbations for the assumed DE background without specifying a microscopic model for CPL or its modified versions.
	 Convergence of the MCMC chains is monitored using the Gelman–Rubin statistic~\cite{gelman_inference_1992}, requiring $R - 1 < 0.015$ for all parameters. Posterior distributions and contour plots are analyzed and visualized using \texttt{GetDist}~\cite{Lewis:2019xzd}.
	
	All models considered have the following eight free parameters with uniform priors: the physical baryon density $\Omega_{\rm b}h^2 \in [0.005, 0.1]$, the physical cold dark matter density $\Omega_{\rm c}h^2 \in [0.001, 0.99]$, the approximate angular scale of the sound horizon at recombination $\theta_{\rm MC} \in [0.5, 10]$, the amplitude of primordial scalar perturbations $\log\qty(10^{10}A_{\rm s}) \in [1.61, 3.91]$, the spectral index of primordial scalar perturbations $n_{\rm s} \in [0.8, 1.2]$, the reionization optical depth $\tau \in [0.01, 0.8]$, and the CPL parameters $w_0 \in [-3, 1]$ and $w_a \in [-3, 2]$.\footnote{The default implementation of the CPL parametrization in \texttt{CAMB} additionally imposes the condition $w_0 + w_a < 0$, which we have also enforced for the modified models.}
	The prior ranges on $w_0$ and $w_a$ follow those adopted in the DESI CPL analyses~\cite{DESI:2024mwx,DESI:2025zgx}. We retain these ranges for all three parametrizations to compare their fits within the same conventional parameter domain.
	
	The datasets used are as follows:
	\begin{itemize}
		\item \textbf{\texttt{CMB}:}\hspace{0.5em}\textit{CMB anisotropies and lensing.} We use the full-mission \textit{Planck} 2018 legacy data release~\cite{Planck:2018nkj,Planck:2019nip}, including the high-$\ell$ TT, TE, and EE likelihoods from \texttt{Plik}, the low-$\ell$ TT likelihood from \texttt{Commander}, and the low-$\ell$ EE likelihood from \texttt{SimAll}. In addition, we include the \texttt{actplanck\_baseline v1.2} likelihood, which combines CMB lensing measurements from the \textit{Planck} PR4 and Atacama Cosmology Telescope (ACT) DR6 datasets~\cite{ACT:2023kun,ACT:2023dou,Carron:2022eyg}.
		
		\item \textbf{\texttt{DESI}:}\hspace{0.5em}\textit{BAO measurements.} We use the latest compilation of BAO distance measurements from DESI DR2, consisting of 16 $D_M/r_{\rm d}$ and $D_H/r_{\rm d}$ measurements across the redshift range $0.4 < z < 4.2$, as well as a $D_V/r_{\rm d}$ measurement at lower redshifts ($0.1 < z < 0.4$), as presented in Table IV of Ref.~\cite{DESI:2025zgx}. These measurements are derived from multiple tracers including ELGs, LRGs, QSOs, and Ly$\alpha$ forests.
		
		\item \textbf{\texttt{Pantheon\textsuperscript{+}}:}\hspace{0.5em}\textit{SNe Ia distance moduli.} We use the Pantheon+ compilation~\cite{Brout:2022vxf}, which includes 1701 light curves of 1550 unique SNeIa in the redshift range $z \in [0.001, 2.26]$, incorporating improved calibration and systematics treatment. When combined with the SH0ES Cepheid host distances (denoted as \texttt{R22+Pantheon\textsuperscript{+}}), the absolute magnitude $M_B$ and hence $H_0$ can also be constrained through the formalism detailed in Eq.~(14) of Ref.~\cite{Brout:2022vxf}.
		
		\item \textbf{\texttt{UNION3}:}\hspace{0.5em}\textit{SNe Ia distance moduli.} We use the UNION3 compilation~\cite{Rubin:2023ovl}, which consists of 2087 SNe Ia from a reanalysis of the original UNION2.1 dataset. The light curves have been homogenized and recalibrated using consistent SALT2 fitting procedures, providing a legacy sample over a broad redshift range with updated systematics and light-curve standardization.
		
		\item \textbf{\texttt{DESY5}:}\hspace{0.5em}\textit{SNe Ia distance moduli.} We use the DESY5 dataset~\cite{DES:2024hip,DES:2024jxu,DES:2024upw}, based on the full five-year data release of the Dark Energy Survey Supernova Program. It contains 1635 spectroscopically confirmed and photometrically classified SNe Ia spanning the redshift range $0.1 < z < 1.13$, and includes complementary low-$z$ samples from Foundation and CSP.
		
		\item \textbf{\texttt{R22}:}\hspace{0.5em}\textit{Local Cepheid calibration of SNe Ia.} We use the $H_0$ prior derived by the SH0ES collaboration in Ref.~\cite{Riess:2021jrx}, based on Cepheid-calibrated distance measurements of SNe Ia in nearby galaxies. This Gaussian prior on $H_0 = 73.04 \pm 1.04$~km s${}^{-1}$ Mpc${}^{-1}$ serves as an external calibration of the cosmic distance ladder and is used in place of BAO data when combined with \texttt{DESY5} or \texttt{UNION3}. In the case of \texttt{Pantheon\textsuperscript{+}}, although we label the combination as \texttt{R22+Pantheon\textsuperscript{+}} for consistency with other naming conventions, we emphasize that we do not apply an external $H_0$ prior. Instead, we use the full \texttt{sn.pantheonplusshoes} likelihood provided in the GitHub page for \texttt{Cobaya},\footnote{\url{https://github.com/CobayaSampler}.} which already incorporates the Cepheid calibration self-consistently.
	\end{itemize}

	\subsection{Results for CPL}
	The central results for the CPL model in our analyses are presented in~\cref{fig:DESI-like}, summarizing the constraints from our data analyses in the $w_0$–$w_a$ parameter plane and illustrating their relation to a PDL crossing. Our data set combinations consist of two main groups: one combining our CMB data set with the DESI DR2 BAO data along with three different SNe Ia compilations that match those employed in Refs.~\cite{DESI:2024mwx,DESI:2025zgx}, and the second replacing the BAO data with R22, i.e., the SH0ES calibration of the SNe Ia. The left panel of~\cref{fig:DESI-like} shows the two-dimensional marginalized posterior distributions for the CPL parameters, $w_0$ and $w_a$. We observe that, for all six data set combinations considered here, the $68\%$ and $95\%$ confidence-level (CL) contours are entirely located within the bottom-right white region, corresponding to a PDL crossing from a phantom-like (P-like, $w < -1$) equation of state to a quintessence-like (Q-like, $w > -1$) behavior as the Universe expands. Interestingly, the contours obtained from the data sets that include \texttt{DESI} measurements align closely along a specific phantom crossing line, while those involving the R22 measurement also align clearly, but along a different crossing line. This difference highlights how distinct calibrations of SNe Ia data sets affect the inferred PDL crossing scale. These calibrations can be either direct distance ladder measurements, as done by the SH0ES group, or a cosmological model-dependent approach, where the sound horizon measurements from the CMB calibrate the BAO data, which in turn calibrate the SNe Ia. Despite the different calibrations, evidence for the PDL crossing is present in both cases.
	
	\begin{table}[b]
		\setlength{\tabcolsep}{4pt}
		\centering
		\caption{Summary of the constraints (68\% CL) on the PDL crossing scale factor $a_{\rm c}$, the probability that it lies outside the physical interval $[0,1]$ based on its one-dimensional marginalized posterior, and the corresponding number of standard deviations required to exclude this probability in a standard Gaussian distribution (shown for reference).
		}
		\begin{tabular}{lccc}
			\toprule
			Data set & $a_{\rm c}$  & $P(a_{\rm c} \notin [0,1])$ &  $\sigma$ \\
			\midrule
			\texttt{DESI+CMB+Pantheon\textsuperscript{+}} & $0.730^{+0.052}_{-0.040}$ & $<\!1.5 \times 10^{-3}$ & $>\!3.1$ \\
			\texttt{DESI+CMB+UNION3}                      & $0.689^{+0.036}_{-0.029}$ & $<\!1.7 \times 10^{-5}$ & $>\!4.3$ \\
			\texttt{DESI+CMB+DESY5}                       & $0.704^{+0.042}_{-0.026}$ & $<\!9.7 \times 10^{-5}$ & $>\!3.9$ \\
			\texttt{R22+CMB+Pantheon\textsuperscript{+}} & $0.821^{+0.015}_{-0.018}$ & $<\!3.1 \times 10^{-6}$ & $>\!4.6$ \\
			\texttt{R22+CMB+UNION3}                       & $0.808^{+0.016}_{-0.021}$ & $<\!1.4 \times 10^{-3}$ & $>\!3.2$ \\
			\texttt{R22+CMB+DESY5}                        & $0.796 \pm 0.015$         & $<\!1.9 \times 10^{-7}$                & $>\!5.2$ \\
			\bottomrule
		\end{tabular}\label{tab:ac}
	\end{table}

	In the right panel of~\cref{fig:DESI-like}, we present the one-dimensional marginalized posterior distributions for the phantom crossing scale factor, $a_{\rm c}$. These posteriors are well peaked within the interval $[0,1]$, strongly indicating a preference for a PDL crossing during the expansion history of the Universe. The numerical constraints on $a_{\rm c}$ at the $68\%$ CL are summarized in~\cref{tab:ac}. To quantify the statistical evidence supporting a PDL crossing within the observable expansion history, we integrate the posterior distributions over the physically irrelevant range $a_{\rm c} \notin [0,1]$. This provides an estimate of the probability that a PDL crossing does not occur within the observable redshift range. We present upper bounds on the probability that a crossing does not exist within the observed history, denoted $P(a_{\rm c} \notin [0,1])$, in the second column of~\cref{tab:ac}. These probabilities were computed numerically using the trapezoidal integration rule. To maintain a conservative assessment, we have included the upper-bound error of the trapezoidal rule in a manner that favors the absence of a PDL crossing. Finally, for reference, the third column of~\cref{tab:ac} provides the equivalent number of standard deviations necessary to exclude these probabilities if they arose from a Gaussian distribution. These standard-deviation equivalents contextualize the significance of the inferred PDL crossing within the familiar framework used for tensions in cosmology. 
	
	Table~\ref{tab:ac} demonstrates that the phantom crossing scale factor, $a_{\rm c}$, is well constrained across all dataset combinations, with posterior means significantly below the upper boundary value $a_{\rm c} = 1$. This leads to very low probabilities that the crossing does not occur within the observable expansion history: about $0.1\%$ in the least constraining case (\texttt{R22+CMB+UNION3}) and below $10^{-5}\%$ in the most constraining case (\texttt{R22+CMB+DESY5}). These probabilities correspond to Gaussian significances ranging from $3.1\sigma$ to $5.2\sigma$, indicating a very strong statistical evidence in favor of a PDL crossing. When \texttt{R22} is used instead of \texttt{DESI}, the constraints on $a_{\rm c}$ tighten considerably, although the central values shift closer to 1. Conversely, dataset combinations including \texttt{DESI} tend to yield lower central values of $a_{\rm c}$, but with significantly larger error bars, by a factor of 2–3, which reduces the overall statistical significance of the crossing compared to the \texttt{R22}-based results. It is important to note that, for both data set groups containing either \texttt{DESI} or \texttt{R22}, the constraints on $a_\mathrm{c}$ agree very well within different choices of SNe Ia compilations as can be seen from~\cref{fig:DESI-like} and~\cref{tab:ac}.

	We take the DESI-based combinations as our main results. The combinations involving \texttt{R22} provide a complementary comparison of the crossing inferred with a different calibration of the SNe Ia distances. However, the values of $H_0$ in Table~\ref{tab:chi2} remain below the SH0ES determination, indicating residual tension between the local calibration and CMB+SNeIa within the models considered. The tighter constraints on $a_{\rm c}$ obtained with \texttt{R22} should therefore be interpreted in the context of this tension, rather than as stronger evidence for a PDL crossing.

	\subsection{Results for the modified CPL models}
	\begin{table}[b]
		\setlength{\tabcolsep}{1pt}
		\centering
		\caption{Comparison of best-fit sample $\chi^2$ values and constraints ($68\%$ CL) on $H_0$ for CPL and its modified versions, $\mathrm{CPL}_{>a_c}$ and $\mathrm{CPL}_{<a_c}$, across different dataset combinations. The standard CPL model corresponds to the baseline $\Delta\chi^2 = 0$, and a positive value indicates a worse fit for the compared model.}
		\resizebox{\linewidth}{!}{%
			\begin{tabular}{ll|c|c|c}
				\toprule
				& & CPL & $\mathrm{CPL}_{>a_c}$ & $\mathrm{CPL}_{<a_c}$ \\
				\midrule
				\multirow{2}{*}{\texttt{DESI+CMB+Pantheon\textsuperscript{+}}} 
				& $\Delta \chi^2$ & 0 & 5.1 & 1.4 \\
				& $H_0$ & $67.64\pm 0.59$ & $67.57\pm 0.56$ & $68.72\pm 0.38$ \\
				\midrule
				\multirow{2}{*}{\texttt{DESI+CMB+DESY5}} 
				& $\Delta \chi^2$ & 0 & 4.7 & 13.3 \\
				& $H_0$ & $66.88\pm 0.56$ & $66.44\pm 0.55$ & $68.57\pm 0.37$ \\
				\midrule
				\multirow{2}{*}{\texttt{DESI+CMB+UNION3}} 
				& $\Delta \chi^2$ & 0 & 8.7 & 10.7 \\
				& $H_0$ & $66.04\pm 0.84$ & $65.93^{+0.82}_{-1.0}$ & $68.73\pm 0.38$ \\
				\midrule
				\multirow{2}{*}{\texttt{R22+CMB+Pantheon\textsuperscript{+}}} 
				& $\Delta \chi^2$ & 0 & 15.3 & 3.5 \\
				& $H_0$ & $70.85\pm 0.72$ & $69.8^{+1.4}_{-1.7}$ & $71.28\pm 0.77$ \\
				\midrule
				\multirow{2}{*}{\texttt{R22+CMB+DESY5}} 
				& $\Delta \chi^2$ & 0 & 18.7 & 9.2 \\
				& $H_0$ & $69.95\pm 0.68$ & $67.80\pm 0.58$ & $69.81\pm 0.79$ \\
				\midrule
				\multirow{2}{*}{\texttt{R22+CMB+UNION3}} 
				& $\Delta \chi^2$ & 0 & 9.2 & 1.4 \\
				& $H_0$ & $70.54\pm 0.79$ & $70.1^{+1.3}_{-0.85}$ & $71.02\pm 0.85$ \\
				\bottomrule
			\end{tabular}%
		}
		\label{tab:chi2}
	\end{table}
	
	We analyze the two modified CPL models introduced in this paper, namely $\mathrm{CPL}_{>a_c}$ and $\mathrm{CPL}_{<a_c}$, for the same six data set combinations used in the analyses of CPL. As shown in~\cref{tab:chi2}, the standard CPL model achieves the lowest sampled $\chi^2$ in all six cases. All three models have the same number of free parameters, so this improvement does not result from adding parameters to the CPL fit. Within the parametrizations and prior ranges considered here, the comparison favors the full CPL evolution over the two modifications that remove the crossing.
	
	The inferred $H_0$ values from CPL and $\mathrm{CPL}_{>a_c}$ are generally consistent, differing by less than $1\sigma$ in all cases except for \texttt{R22+CMB+DESY5}, where $\mathrm{CPL}_{>a_c}$ yields a slightly lower value. The $\mathrm{CPL}_{<a_c}$ model tends to produce higher $H_0$ values. However, even in these cases, the resulting $H_0$ values do not fully resolve the Hubble tension. When \texttt{R22} is not included in the dataset combination, comparisons with the \texttt{R22} measurement of $H_0 = 73.04 \pm 1.04$~km\,s${}^{-1}$\,Mpc${}^{-1}$~\cite{Riess:2021jrx} result in a tension of at least $\sim\!4\sigma$. Including \texttt{R22} in the dataset combination, as expected, moves the Hubble constant toward the local value, reaching its highest value in the case of \texttt{R22+CMB+Pantheon\textsuperscript{+}}, for which the constraint is $H_0 = 71.28 \pm 0.77$~km\,s${}^{-1}$\,Mpc${}^{-1}$. 
	
\begin{figure}[t]
		\centering    \includegraphics[width=0.48\textwidth]{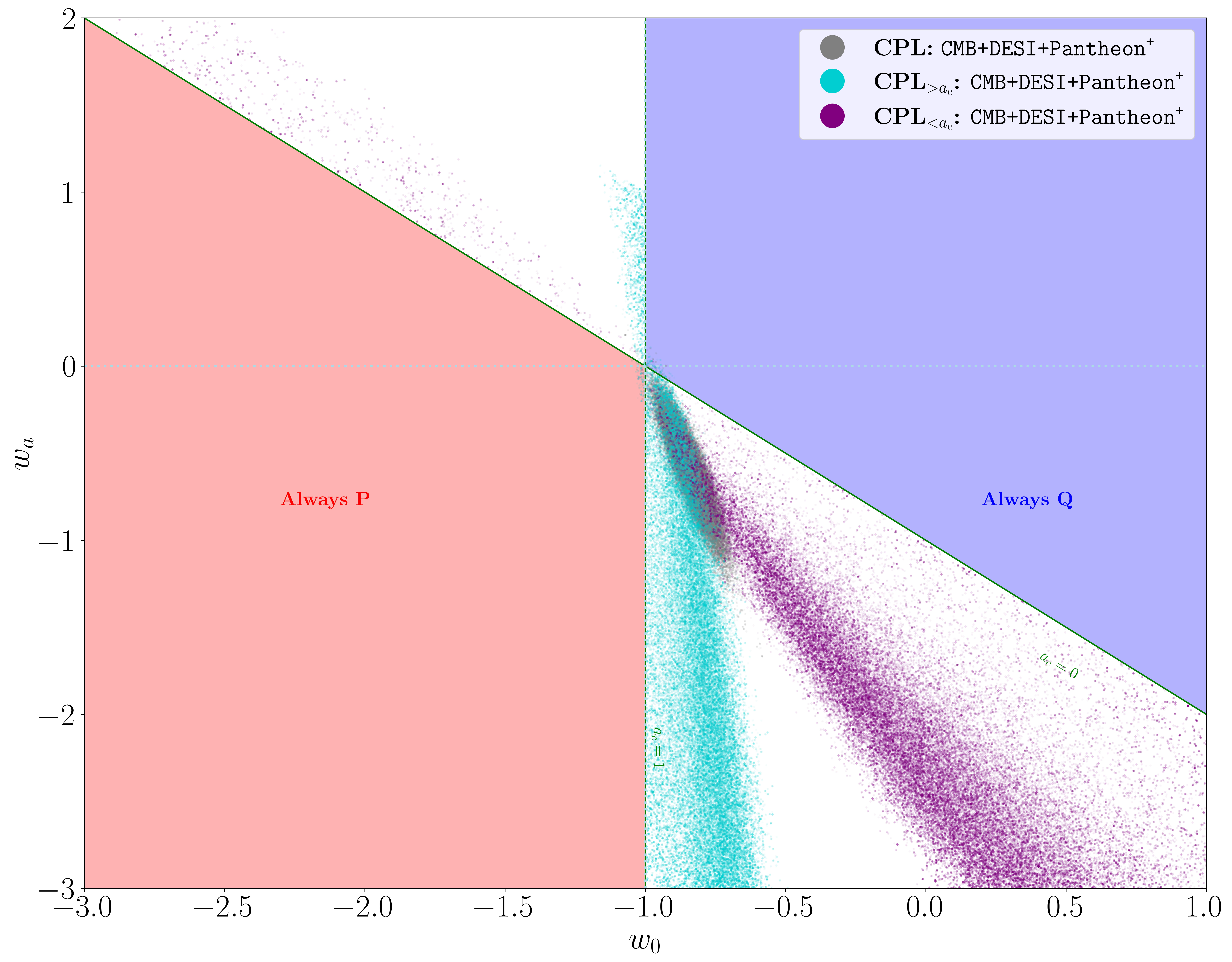}    \includegraphics[width=0.48\textwidth]{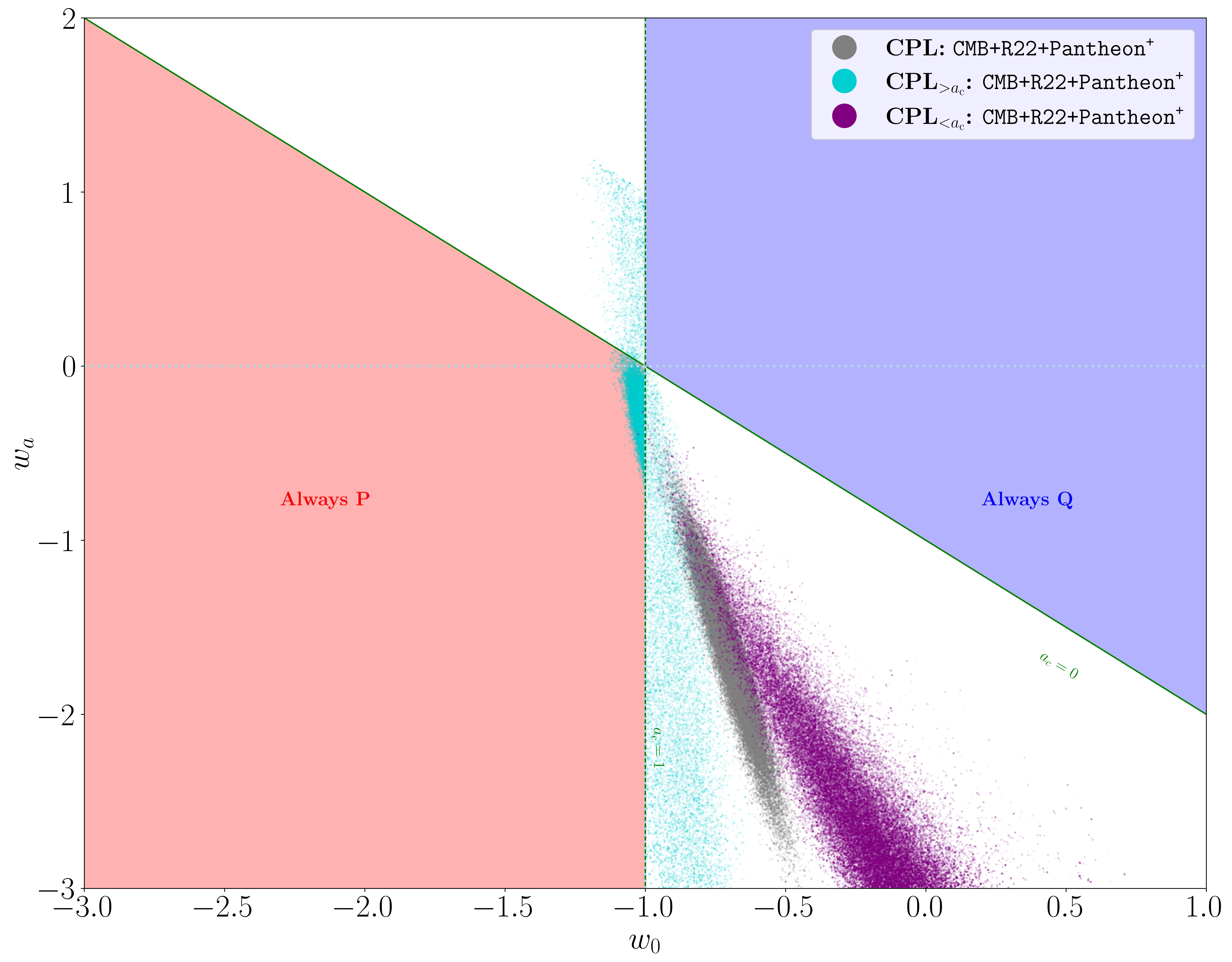}
		\caption{Comparison of the modified models and standard CPL in the $w_0$--$w_a$ plane using the \texttt{Pantheon\textsuperscript{+}} dataset. We plot the full MCMC chains to preserve sharp features of the posteriors, and scale point transparency by the weight of the chains. Darker regions indicate higher posterior probability. Results are qualitatively similar for other SNe~Ia datasets.}
		\label{fig:divided}
	\end{figure}

	We note that the posterior distributions of $w_0$ and $w_a$ in the modified models reach the boundaries of the prior ranges (see~\cref{fig:divided}). However, for both modified models and all data combinations, the lowest-$\chi^2$ sampled points lie inside the adopted $w_0$--$w_a$ prior ranges. The distinction is particularly clear for $\mathrm{CPL}_{>a_c}$, whose best samples are separated from the $w_a=-3$ boundary. Some best samples of $\mathrm{CPL}_{<a_c}$ lie relatively close to this boundary, so extending the priors could alter the precise numerical fit differences. Nevertheless, this has limited impact on our main conclusions, as the crossing probabilities are obtained independently from the standard CPL analysis. Moreover, the differences of $\Delta\chi^2=1.4$ for $\mathrm{CPL}_{<a_c}$ in the \texttt{DESI+CMB+Pantheon\textsuperscript{+}} and \texttt{R22+CMB+UNION3} cases already indicate comparable fits rather than a decisive preference for either model. We therefore interpret~\cref{tab:chi2} as a comparison within the stated prior ranges. \footnote{The $\chi^2$ values reported in Table~\ref{tab:chi2} are obtained from the lowest-$\chi^2$ samples in the MCMC chains, without dedicated minimizer. Consequently, small differences in $\Delta\chi^2$ should be interpreted with caution.}
		
	Finally, in~\cref{fig:divided}, we show the results for the modified models in comparison with the standard CPL parametrization within the $w_0$--$w_a$ plane. To maintain clarity in the figure, we include results from only one SNe Ia dataset, choosing \texttt{Pantheon\textsuperscript{+}} due to its robust integration with the SH0ES calibration; however, the results are qualitatively similar for the other SNe Ia datasets. Note that we directly plot the full MCMC chains instead of contours, as sharp features in the posteriors are otherwise lost during the smoothing processes involved in contour construction. The transparency of the points is scaled according to their weights in the MCMC chains. Darker regions, caused by the density of the chains, correspond to higher posterior probability.
	The top panel clearly demonstrates that the standard CPL model lies at the intersection of the modified models $\mathrm{CPL}_{>a_c}$ and $\mathrm{CPL}_{<a_c}$. This indicates that even when the DE EoS is forced to behave as a cosmological constant during part of its evolution, the remaining portion described by the CPL parametrization remains consistent with the constraints obtained from the standard CPL parametrization describing the entire evolution history. In contrast, as shown in the bottom panel, this consistency is lost when \texttt{DESI} is replaced by \texttt{R22}; the modified models favor different evolution histories compared to CPL, even within the regions where their functional forms coincide.
	In both panels, $\mathrm{CPL}_{<a_c}$ is almost entirely contained within the white regions that correspond to would-be PDL crossings not allowed by this model. In the top-left white region, $\mathrm{CPL}_{<a_c}$ exhibits an initially Q-like behavior that then freezes into a constant energy density at $a_{\rm c}$, while in the bottom-right white region, it exhibits an initially P-like behavior that similarly freezes into a constant energy density. As seen in the figure, the large majority of the posterior probability for this model lies in the bottom-right region, indicating a preference for an initially P-like equation of state.
	Moving on to the $\mathrm{CPL}_{>a_c}$ model, the top-left white region corresponds to an initially constant energy density that thaws into a P-like behavior, while the bottom-right white region corresponds to an initially constant energy density that thaws into a Q-like behavior. In the top panel, which includes \texttt{DESI}, most of the posterior probability lies in the bottom-right region, indicating a preference for a late-time Q-like behavior. However, in the bottom panel, when \texttt{DESI} is replaced by \texttt{R22}, the highest-density part of the posterior shifts into the red region corresponding to a fully phantom equation of state throughout the expansion history. In this region, the model becomes indistinguishable from CPL in its functional form, as the modification has no effect when $a_{\rm c} \notin [0,1]$, yet, the vanilla CPL model does not prefer this region for any of our data set combinations.

	\section{Conclusions} \label{sec:Conclusion}
	In this work, we have investigated the observational evidence for a PDL crossing in the DE EoS parameter using the CPL parametrization and two physically motivated modifications, $\mathrm{CPL}_{>a_c}$ and $\mathrm{CPL}_{<a_c}$, which forbid such a crossing by construction. By analyzing a wide range of dataset combinations, including CMB data, DESI DR2 BAO, and three SNe Ia compilations, we find strong statistical support for a crossing scenario occurring within the observable expansion history of the Universe.
	
	Our constraints on the crossing scale factor $a_{\rm c}$ show that it is consistently located well within the physical range $[0,1]$, with exclusion probabilities for $a_{\rm c} \notin [0,1]$ as low as $10^{-5}\%$ in the most constraining cases. This translates into statistical significances ranging from $3.1\sigma$ to $5.2\sigma$, providing compelling evidence in favor of a PDL crossing. Moreover, the standard CPL model achieves lower sampled $\chi^2$ values than both non-crossing modifications across all dataset combinations within the common prior ranges. This comparison supports the preference for the full CPL evolution relative to these two alternatives in the parameter domain explored.
	
	We find that $\mathrm{CPL}_{<a_c}$ generally yields higher values of $H_0$ but not an amount adequate for a convincing resolution of the Hubble tension, whereas $\mathrm{CPL}_{>a_c}$ yields $H_0$ values more consistent with the standard CPL model. Interestingly, in the $w_0$–$w_a$ plane, the standard CPL contours lie at the intersection of the two modified models for our data set combinations including \texttt{DESI}, suggesting that the full CPL parametrization flexibly captures both early and late-time dynamics required by the data. However, when \texttt{DESI} is replaced by \texttt{R22}, this consistency is lost, and the modified models trace out distinct, incompatible trajectories. The relative fits show that the full CPL form accommodates these data combinations more successfully than the two modifications within the adopted priors.
	
	The \texttt{DESI}-based combinations provide our main results, with a significance of at least $3.1\sigma$ for a PDL crossing ($a_{\rm c}\in[0,1]$). Although the combinations involving \texttt{R22} also favor a crossing and yield tighter constraints on $a_{\rm c}$, the inferred $H_0$ values remain below the SH0ES determination, indicating residual tension between the local calibration and the CMB+SNeIa data. These combinations illustrate the effect of the local calibration on the inferred crossing, but their tighter constraints should not be interpreted as more robust evidence for it.

	Several studies in the literature have raised concerns that the CPL analysis of current data may be misleading, suggesting that a thawing quintessence model could suffice to describe the observed behavior of dark energy and that the apparent evidence for a PDL crossing within CPL may be an artifact~\cite{Shlivko:2024llw,Wolf:2024eph,Notari:2024rti,DESI:2025fii}. Within the conventional CPL prior ranges, our results show that $\mathrm{CPL}_{>a_c}$, which phenomenologically mimics thawing dark energy without allowing a PDL crossing, yields a higher sampled $\chi^2$ than CPL in every combination, with $\Delta\chi^2$ ranging from 4.7 to 18.7. The lowest-$\chi^2$ samples lie well within the prior ranges.
	 Our analysis of $\mathrm{CPL}_{<a_c}$, which transitions from a dynamical EoS parameter to a cosmological constant, shows that this model can occasionally yield a fit comparable to CPL, with $\Delta\chi^2$ differences as low as 1.4 in some data combinations. 

	 The fit comparison and the crossing probabilities inferred within standard CPL provide distinct information. The latter are obtained from the standard CPL posterior and do not depend on the sampling of the modified models. The former assesses the two non-crossing modifications within the same prior ranges adopted for CPL. Extending these ranges could alter the precise numerical $\Delta\chi^2$ values, particularly in cases where the models already yield comparable fits.
		
	In summary, our findings offer strong empirical support for a PDL crossing in the dark energy sector and highlight the utility of testing restricted models to isolate physical features. Moreover, Ref.~\cite{Wang:2024dka} shows that combining CPL with early dark energy can alleviate the Hubble tension and reduce the evidence for a PDL crossing. Exploring combinations of CPL with early-time modifications could represent a promising scenario—particularly a combination with a varying electron mass, which is compelling as it could allow the total number of extra free parameters to remain at three.  High-precision surveys will be essential for confirming these results and for further probing the dynamics of dark energy beyond simple parameterizations.\looseness=-1
	
	\begin{acknowledgments}
		\noindent EDV is supported by a Royal Society Dorothy Hodgkin Research Fellowship. WG is supported from National Aeronautics and Space Administration (NASA) under Grant No.~80NSSC24K0898. 
		We acknowledge the IT Services at The University of Sheffield for the provision of services for High Performance Computing. 
		This article is based upon work from the COST Action CA21136 - ``Addressing observational tensions in cosmology with systematics and fundamental physics (CosmoVerse)'', supported by COST - ``European Cooperation in Science and Technology''.
	\end{acknowledgments}

\bibliography{main}

@article{Abdalla:2022yfr,
    author = "Abdalla, Elcio and others",
    title = "{Cosmology intertwined: A review of the particle physics, astrophysics, and cosmology associated with the cosmological tensions and anomalies}",
    eprint = "2203.06142",
    archivePrefix = "arXiv",
    primaryClass = "astro-ph.CO",
    reportNumber = "FERMILAB-CONF-22-192-SCD",
    doi = "10.1016/j.jheap.2022.04.002",
    journal = "JHEAp",
    volume = "34",
    pages = "49--211",
    year = "2022"
}

@article{Perivolaropoulos:2021jda,
    author = "Perivolaropoulos, Leandros and Skara, Foteini",
    title = "{Challenges for {\ensuremath{\Lambda}}CDM: An update}",
    eprint = "2105.05208",
    archivePrefix = "arXiv",
    primaryClass = "astro-ph.CO",
    doi = "10.1016/j.newar.2022.101659",
    journal = "New Astron. Rev.",
    volume = "95",
    pages = "101659",
    year = "2022"
}

@article{DiValentino:2022fjm,
    author = "Di Valentino, Eleonora",
    title = "{Challenges of the Standard Cosmological Model}",
    doi = "10.3390/universe8080399",
    journal = "Universe",
    volume = "8",
    number = "8",
    pages = "399",
    year = "2022"
}

@article{CosmoVerse:2025txj,
    author = "Di Valentino, Eleonora and others",
    collaboration = "CosmoVerse Network",
    title = "{The CosmoVerse White Paper: Addressing observational tensions in cosmology with systematics and fundamental physics}",
    eprint = "2504.01669",
    archivePrefix = "arXiv",
    primaryClass = "astro-ph.CO",
    doi = "10.1016/j.dark.2025.101965",
    journal = "Phys. Dark Univ.",
    volume = "49",
    pages = "101965",
    year = "2025"
}

@article{Verde:2019ivm,
    author = "Verde, L. and Treu, T. and Riess, A. G.",
    title = "{Tensions between the Early and the Late Universe}",
    eprint = "1907.10625",
    archivePrefix = "arXiv",
    primaryClass = "astro-ph.CO",
    doi = "10.1038/s41550-019-0902-0",
    journal = "Nature Astron.",
    volume = "3",
    pages = "891",
    year = "2019"
}

@article{DiValentino:2020zio,
    author = "Di Valentino, Eleonora and others",
    title = "{Snowmass2021 - Letter of interest cosmology intertwined II: The hubble constant tension}",
    eprint = "2008.11284",
    archivePrefix = "arXiv",
    primaryClass = "astro-ph.CO",
    reportNumber = "FERMILAB-PUB-21-590-PPD",
    doi = "10.1016/j.astropartphys.2021.102605",
    journal = "Astropart. Phys.",
    volume = "131",
    pages = "102605",
    year = "2021"
}

@article{DiValentino:2021izs,
    author = "Di Valentino, Eleonora and Mena, Olga and Pan, Supriya and Visinelli, Luca and Yang, Weiqiang and Melchiorri, Alessandro and Mota, David F. and Riess, Adam G. and Silk, Joseph",
    title = "{In the realm of the Hubble tension{\textemdash}a review of solutions}",
    eprint = "2103.01183",
    archivePrefix = "arXiv",
    primaryClass = "astro-ph.CO",
    reportNumber = "IPPP/20/108",
    doi = "10.1088/1361-6382/ac086d",
    journal = "Class. Quant. Grav.",
    volume = "38",
    number = "15",
    pages = "153001",
    year = "2021"
}

@article{Schoneberg:2021qvd,
    author = {Sch{\"o}neberg, Nils and Franco Abell{\'a}n, Guillermo and P{\'e}rez S{\'a}nchez, Andrea and Witte, Samuel J. and Poulin, Vivian and Lesgourgues, Julien},
    title = "{The H0 Olympics: A fair ranking of proposed models}",
    eprint = "2107.10291",
    archivePrefix = "arXiv",
    primaryClass = "astro-ph.CO",
    doi = "10.1016/j.physrep.2022.07.001",
    journal = "Phys. Rept.",
    volume = "984",
    pages = "1--55",
    year = "2022"
}

@article{Shah:2021onj,
    author = "Shah, Paul and Lemos, Pablo and Lahav, Ofer",
    title = "{A buyer{\textquoteright}s guide to the Hubble constant}",
    eprint = "2109.01161",
    archivePrefix = "arXiv",
    primaryClass = "astro-ph.CO",
    doi = "10.1007/s00159-021-00137-4",
    journal = "Astron. Astrophys. Rev.",
    volume = "29",
    number = "1",
    pages = "9",
    year = "2021"
}

@article{Kamionkowski:2022pkx,
    author = "Kamionkowski, Marc and Riess, Adam G.",
    title = "{The Hubble Tension and Early Dark Energy}",
    eprint = "2211.04492",
    archivePrefix = "arXiv",
    primaryClass = "astro-ph.CO",
    doi = "10.1146/annurev-nucl-111422-024107",
    journal = "Ann. Rev. Nucl. Part. Sci.",
    volume = "73",
    pages = "153--180",
    year = "2023"
}

@article{Giare:2023xoc,
    author = "Giar{\`e}, William",
    title = "{CMB Anomalies and the Hubble Tension}",
    eprint = "2305.16919",
    archivePrefix = "arXiv",
    primaryClass = "astro-ph.CO",
    doi = "10.1007/978-981-99-0177-7\_36",
    month = "5",
    year = "2023"
}

@article{Hu:2023jqc,
    author = "Hu, Jian-Ping and Wang, Fa-Yin",
    title = "{Hubble Tension: The Evidence of New Physics}",
    eprint = "2302.05709",
    archivePrefix = "arXiv",
    primaryClass = "astro-ph.CO",
    doi = "10.3390/universe9020094",
    journal = "Universe",
    volume = "9",
    number = "2",
    pages = "94",
    year = "2023"
}

@article{Verde:2023lmm,
    author = {Verde, Licia and Sch{\"o}neberg, Nils and Gil-Mar{\'\i}n, H{\'e}ctor},
    title = "{A Tale of Many H0}",
    eprint = "2311.13305",
    archivePrefix = "arXiv",
    primaryClass = "astro-ph.CO",
    doi = "10.1146/annurev-astro-052622-033813",
    journal = "Ann. Rev. Astron. Astrophys.",
    volume = "62",
    number = "1",
    pages = "287--331",
    year = "2024"
}

@book{DiValentino:2024yew,
    editor = "Di Valentino, Eleonora and Brout, Dillon",
    title = "{The Hubble Constant Tension}",
    doi = "10.1007/978-981-99-0177-7",
    isbn = "978-981-99-0176-0, 978-981-99-0179-1, 978-981-99-0177-7",
    publisher = "Springer",
    series = "Springer Series in Astrophysics and Cosmology",
    year = "2024"
}

@article{Li:2026xaz,
    author = "Li, Tian-Nuo and Giar{\`e}, William and Du, Guo-Hong and Li, Yun-He and Di Valentino, Eleonora and Zhang, Jing-Fei and Zhang, Xin",
    title = "{Robust Preference for Dark Sector Interactions}",
    eprint = "2601.07361",
    archivePrefix = "arXiv",
    primaryClass = "astro-ph.CO",
    month = "1",
    year = "2026"
}

@article{Riess:2021jrx,
    author = "Riess, Adam G. and others",
    title = "{A Comprehensive Measurement of the Local Value of the Hubble Constant with 1 km s$^{−1}$ Mpc$^{−1}$ Uncertainty from the Hubble Space Telescope and the SH0ES Team}",
    eprint = "2112.04510",
    archivePrefix = "arXiv",
    primaryClass = "astro-ph.CO",
    doi = "10.3847/2041-8213/ac5c5b",
    journal = "Astrophys. J. Lett.",
    volume = "934",
    number = "1",
    pages = "L7",
    year = "2022"
}

@article{Murakami:2023xuy,
    author = "Murakami, Yukei S. and Riess, Adam G. and Stahl, Benjamin E. and Kenworthy, W. D'Arcy and Pluck, Dahne-More A. and Macoretta, Antonella and Brout, Dillon and Jones, David O. and Scolnic, Dan M. and Filippenko, Alexei V.",
    title = "{Leveraging SN Ia spectroscopic similarity to improve the measurement of H $_{0}$}",
    eprint = "2306.00070",
    archivePrefix = "arXiv",
    primaryClass = "astro-ph.CO",
    doi = "10.1088/1475-7516/2023/11/046",
    journal = "JCAP",
    volume = "11",
    pages = "046",
    year = "2023"
}

@article{Breuval:2024lsv,
    author = "Breuval, Louise and Riess, Adam G. and Casertano, Stefano and Yuan, Wenlong and Macri, Lucas M. and Romaniello, Martino and Murakami, Yukei S. and Scolnic, Daniel and Anand, Gagandeep S. and Soszy{\'n}ski, Igor",
    title = "{Small Magellanic Cloud Cepheids Observed with the Hubble Space Telescope Provide a New Anchor for the SH0ES Distance Ladder}",
    eprint = "2404.08038",
    archivePrefix = "arXiv",
    primaryClass = "astro-ph.CO",
    doi = "10.3847/1538-4357/ad630e",
    journal = "Astrophys. J.",
    volume = "973",
    number = "1",
    pages = "30",
    year = "2024"
}

@article{Planck:2018vyg,
    author = "Aghanim, N. and others",
    collaboration = "Planck",
    title = "{Planck 2018 results. VI. Cosmological parameters}",
    eprint = "1807.06209",
    archivePrefix = "arXiv",
    primaryClass = "astro-ph.CO",
    doi = "10.1051/0004-6361/201833910",
    journal = "Astron. Astrophys.",
    volume = "641",
    pages = "A6",
    year = "2020",
    note = "[Erratum: Astron.Astrophys. 652, C4 (2021)]"
}

@article{ACT:2025fju,
    author = "Louis, Thibaut and others",
    collaboration = "Atacama Cosmology Telescope",
    title = "{The Atacama Cosmology Telescope: DR6 power spectra, likelihoods and {\ensuremath{\Lambda}}CDM parameters}",
    eprint = "2503.14452",
    archivePrefix = "arXiv",
    primaryClass = "astro-ph.CO",
    reportNumber = "FERMILAB-PUB-25-0071-PPD",
    doi = "10.1088/1475-7516/2025/11/062",
    journal = "JCAP",
    volume = "11",
    pages = "062",
    year = "2025"
}

@article{SPT:2023jql,
    author = "Pan, Z. and others",
    collaboration = "SPT",
    title = "{Measurement of gravitational lensing of the cosmic microwave background using SPT-3G 2018 data}",
    eprint = "2308.11608",
    archivePrefix = "arXiv",
    primaryClass = "astro-ph.CO",
    doi = "10.1103/PhysRevD.108.122005",
    journal = "Phys. Rev. D",
    volume = "108",
    number = "12",
    pages = "122005",
    year = "2023"
}

@article{Freedman:2020dne,
    author = "Freedman, Wendy L. and Madore, Barry F. and Hoyt, Taylor and Jang, In Sung and Beaton, Rachael and Lee, Myung Gyoon and Monson, Andrew and Neeley, Jill and Rich, Jeffrey",
    title = "{Calibration of the Tip of the Red Giant Branch}",
    eprint = "2002.01550",
    archivePrefix = "arXiv",
    primaryClass = "astro-ph.GA",
    doi = "10.3847/1538-4357/ab7339",
    journal = "Astrophys. J.",
    volume = "891",
    number = "1",
    pages = "57",
    year = "2020"
}

@article{Birrer:2020tax,
    author = "Birrer, S. and others",
    title = "{TDCOSMO - IV. Hierarchical time-delay cosmography {\textendash} joint inference of the Hubble constant and galaxy density profiles}",
    eprint = "2007.02941",
    archivePrefix = "arXiv",
    primaryClass = "astro-ph.CO",
    doi = "10.1051/0004-6361/202038861",
    journal = "Astron. Astrophys.",
    volume = "643",
    pages = "A165",
    year = "2020"
}

@article{Anderson:2023aga,
    author = "Anderson, Richard I. and Koblischke, Nolan W. and Eyer, Laurent",
    title = "{Small-amplitude Red Giants Elucidate the Nature of the Tip of the Red Giant Branch as a Standard Candle}",
    eprint = "2303.04790",
    archivePrefix = "arXiv",
    primaryClass = "astro-ph.CO",
    doi = "10.3847/2041-8213/ad284d",
    journal = "Astrophys. J. Lett.",
    volume = "963",
    number = "2",
    pages = "L43",
    year = "2024"
}

@article{Scolnic:2023mrv,
    author = "Scolnic, D. and Riess, A. G. and Wu, J. and Li, S. and Anand, G. S. and Beaton, R. and Casertano, S. and Anderson, R. I. and Dhawan, S. and Ke, X.",
    title = "{CATS: The Hubble Constant from Standardized TRGB and Type Ia Supernova Measurements}",
    eprint = "2304.06693",
    archivePrefix = "arXiv",
    primaryClass = "astro-ph.CO",
    doi = "10.3847/2041-8213/ace978",
    journal = "Astrophys. J. Lett.",
    volume = "954",
    number = "1",
    pages = "L31",
    year = "2023"
}

@article{Jones:2022mvo,
    author = "Jones, D. O. and others",
    title = "{Cosmological Results from the RAISIN Survey: Using Type Ia Supernovae in the Near Infrared as a Novel Path to Measure the Dark Energy Equation of State}",
    eprint = "2201.07801",
    archivePrefix = "arXiv",
    primaryClass = "astro-ph.CO",
    doi = "10.3847/1538-4357/ac755b",
    journal = "Astrophys. J.",
    volume = "933",
    number = "2",
    pages = "172",
    year = "2022"
}

@article{Anand:2021sum,
    author = "Anand, Gagandeep S. and Tully, R. Brent and Rizzi, Luca and Riess, Adam G. and Yuan, Wenlong",
    title = "{Comparing Tip of the Red Giant Branch Distance Scales: An Independent Reduction of the Carnegie-Chicago Hubble Program and the Value of the Hubble Constant}",
    eprint = "2108.00007",
    archivePrefix = "arXiv",
    primaryClass = "astro-ph.CO",
    doi = "10.3847/1538-4357/ac68df",
    journal = "Astrophys. J.",
    volume = "932",
    number = "1",
    pages = "15",
    year = "2022"
}

@article{Freedman:2021ahq,
    author = "Freedman, Wendy L.",
    title = "{Measurements of the Hubble Constant: Tensions in Perspective}",
    eprint = "2106.15656",
    archivePrefix = "arXiv",
    primaryClass = "astro-ph.CO",
    doi = "10.3847/1538-4357/ac0e95",
    journal = "Astrophys. J.",
    volume = "919",
    number = "1",
    pages = "16",
    year = "2021"
}

@article{Uddin:2023iob,
    author = "Uddin, Syed A. and others",
    title = "{Carnegie Supernova Project I and II: Measurements of H $_{0}$ Using Cepheid, Tip of the Red Giant Branch, and Surface Brightness Fluctuation Distance Calibration to Type Ia Supernovae*}",
    eprint = "2308.01875",
    archivePrefix = "arXiv",
    primaryClass = "astro-ph.CO",
    doi = "10.3847/1538-4357/ad3e63",
    journal = "Astrophys. J.",
    volume = "970",
    number = "1",
    pages = "72",
    year = "2024"
}

@article{Huang:2023frr,
    author = "Huang, Caroline D. and others",
    title = "{The Mira Distance to M101 and a 4{\%} Measurement of H $_{0}$}",
    eprint = "2312.08423",
    archivePrefix = "arXiv",
    primaryClass = "astro-ph.CO",
    doi = "10.3847/1538-4357/ad1ff8",
    journal = "Astrophys. J.",
    volume = "963",
    number = "2",
    pages = "83",
    year = "2024"
}

@article{Li:2024yoe,
    author = "Li, Siyang and Riess, Adam G. and Casertano, Stefano and Anand, Gagandeep S. and Scolnic, Daniel M. and Yuan, Wenlong and Breuval, Louise and Huang, Caroline D.",
    title = "{Reconnaissance with JWST of the J-region Asymptotic Giant Branch in Distance Ladder Galaxies: From Irregular Luminosity Functions to Approximation of the Hubble Constant}",
    eprint = "2401.04777",
    archivePrefix = "arXiv",
    primaryClass = "astro-ph.CO",
    doi = "10.3847/1538-4357/ad2f2b",
    journal = "Astrophys. J.",
    volume = "966",
    number = "1",
    pages = "20",
    year = "2024"
}

@article{Pesce:2020xfe,
    author = "Pesce, D. W. and others",
    title = "{The Megamaser Cosmology Project. XIII. Combined Hubble constant constraints}",
    eprint = "2001.09213",
    archivePrefix = "arXiv",
    primaryClass = "astro-ph.CO",
    doi = "10.3847/2041-8213/ab75f0",
    journal = "Astrophys. J. Lett.",
    volume = "891",
    number = "1",
    pages = "L1",
    year = "2020"
}

@article{Kourkchi:2020iyz,
    author = "Kourkchi, Ehsan and Tully, R. Brent and Anand, Gagandeep S. and Courtois, Helene M. and Dupuy, Alexandra and Neill, James D. and Rizzi, Luca and Seibert, Mark",
    title = "{Cosmicflows-4: The Calibration of Optical and Infrared Tully{\textendash}Fisher Relations}",
    eprint = "2004.14499",
    archivePrefix = "arXiv",
    primaryClass = "astro-ph.GA",
    doi = "10.3847/1538-4357/ab901c",
    journal = "Astrophys. J.",
    volume = "896",
    number = "1",
    pages = "3",
    year = "2020"
}

@article{Schombert:2020pxm,
    author = "Schombert, James and McGaugh, Stacy and Lelli, Federico",
    title = "{Using the Baryonic Tully{\textendash}Fisher Relation to Measure H o}",
    eprint = "2006.08615",
    archivePrefix = "arXiv",
    primaryClass = "astro-ph.CO",
    doi = "10.3847/1538-3881/ab9d88",
    journal = "Astron. J.",
    volume = "160",
    number = "2",
    pages = "71",
    year = "2020"
}

@article{Blakeslee:2021rqi,
    author = "Blakeslee, John P. and Jensen, Joseph B. and Ma, Chung-Pei and Milne, Peter A. and Greene, Jenny E.",
    title = "{The Hubble Constant from Infrared Surface Brightness Fluctuation Distances}",
    eprint = "2101.02221",
    archivePrefix = "arXiv",
    primaryClass = "astro-ph.CO",
    doi = "10.3847/1538-4357/abe86a",
    journal = "Astrophys. J.",
    volume = "911",
    number = "1",
    pages = "65",
    year = "2021"
}

@article{deJaeger:2022lit,
    author = "de Jaeger, T. and Galbany, L. and Riess, A. G. and Stahl, B. E. and Shappee, B. J. and Filippenko, A. V. and Zheng, W.",
    title = "{A 5~per{\,}cent measurement of the Hubble{\textendash}Lema{\^\i}tre constant from Type II supernovae}",
    eprint = "2203.08974",
    archivePrefix = "arXiv",
    primaryClass = "astro-ph.CO",
    doi = "10.1093/mnras/stac1661",
    journal = "Mon. Not. Roy. Astron. Soc.",
    volume = "514",
    number = "3",
    pages = "4620--4628",
    year = "2022"
}

@article{Freedman:2024eph,
    author = "Freedman, Wendy L. and Madore, Barry F. and Hoyt, Taylor J. and Jang, In Sung and Lee, Abigail J. and Owens, Kayla A.",
    title = "{Status Report on the Chicago-Carnegie Hubble Program (CCHP): Measurement of the Hubble Constant Using the Hubble and James Webb Space Telescopes}",
    eprint = "2408.06153",
    archivePrefix = "arXiv",
    primaryClass = "astro-ph.CO",
    doi = "10.3847/1538-4357/adce78",
    journal = "Astrophys. J.",
    volume = "985",
    number = "2",
    pages = "203",
    year = "2025",
    note = "[Erratum: Astrophys.J. 993, 252 (2025)]"
}

@article{Riess:2024vfa,
    author = "Riess, Adam G. and others",
    title = "{JWST Validates HST Distance Measurements: Selection of Supernova Subsample Explains Differences in JWST Estimates of Local H $_{0}$}",
    eprint = "2408.11770",
    archivePrefix = "arXiv",
    primaryClass = "astro-ph.CO",
    doi = "10.3847/1538-4357/ad8c21",
    journal = "Astrophys. J.",
    volume = "977",
    number = "1",
    pages = "120",
    year = "2024"
}

@article{Vogl:2024bum,
    author = "Vogl, Christian and others",
    title = "{No rungs attached: A distance-ladder-free determination of the Hubble constant through type II supernova spectral modelling}",
    eprint = "2411.04968",
    archivePrefix = "arXiv",
    primaryClass = "astro-ph.CO",
    doi = "10.1051/0004-6361/202452910",
    journal = "Astron. Astrophys.",
    volume = "702",
    pages = "A41",
    year = "2025"
}

@article{Scolnic:2024hbh,
    author = "Scolnic, Daniel and others",
    title = "{The Hubble Tension in Our Own Backyard: DESI and the Nearness of the Coma Cluster}",
    eprint = "2409.14546",
    archivePrefix = "arXiv",
    primaryClass = "astro-ph.CO",
    doi = "10.3847/2041-8213/ada0bd",
    journal = "Astrophys. J. Lett.",
    volume = "979",
    number = "1",
    pages = "L9",
    year = "2025"
}

@article{Said:2024pwm,
    author = "Said, Khaled and others",
    title = "{DESI peculiar velocity survey {\textendash} Fundamental Plane}",
    eprint = "2408.13842",
    archivePrefix = "arXiv",
    primaryClass = "astro-ph.CO",
    doi = "10.1093/mnras/staf700",
    journal = "Mon. Not. Roy. Astron. Soc.",
    volume = "539",
    number = "4",
    pages = "3627--3644",
    year = "2025"
}

@article{Boubel:2024cqw,
    author = "Boubel, Paula and Colless, Matthew and Said, Khaled and Staveley-Smith, Lister",
    title = "{An improved Tully{\textendash}Fisher estimate of H0}",
    eprint = "2408.03660",
    archivePrefix = "arXiv",
    primaryClass = "astro-ph.CO",
    doi = "10.1093/mnras/stae1925",
    journal = "Mon. Not. Roy. Astron. Soc.",
    volume = "533",
    number = "2",
    pages = "1550--1559",
    year = "2024"
}

@article{Scolnic:2024oth,
    author = "Scolnic, Daniel and Boubel, Paula and Byrne, Jakob and Riess, Adam G. and Anand, Gagandeep S.",
    title = "{Calibrating the Tully-Fisher Relation to Measure the Hubble Constant}",
    eprint = "2412.08449",
    archivePrefix = "arXiv",
    primaryClass = "astro-ph.CO",
    month = "12",
    year = "2024"
}

@article{Li:2025ife,
    author = "Li, Siyang and Riess, Adam G. and Scolnic, Daniel and Casertano, Stefano and Anand, Gagandeep S.",
    title = "{JAGB 2.0: Improved Constraints on the J-region Asymptotic Giant Branch{\textendash}based Hubble Constant from an Expanded Sample of JWST Observations}",
    eprint = "2502.05259",
    archivePrefix = "arXiv",
    primaryClass = "astro-ph.CO",
    doi = "10.3847/1538-4357/addd0c",
    journal = "Astrophys. J.",
    volume = "988",
    number = "1",
    pages = "97",
    year = "2025"
}

@article{Jensen:2025aai,
    author = "Jensen, Joseph B. and Blakeslee, John P. and Cantiello, Michele and Cowles, Mikaela and Anand, Gagandeep S. and Tully, R. Brent and Kourkchi, Ehsan and Raimondo, Gabriella",
    title = "{The TRGB-SBF Project. III. Refining the HST Surface Brightness Fluctuation Distance Scale Calibration with JWST}",
    eprint = "2502.15935",
    archivePrefix = "arXiv",
    primaryClass = "astro-ph.CO",
    doi = "10.3847/1538-4357/addfd6",
    month = "6",
    year = "2025"
}

@article{Riess:2019qba,
    author = "Riess, Adam G.",
    title = "{The Expansion of the Universe is Faster than Expected}",
    eprint = "2001.03624",
    archivePrefix = "arXiv",
    primaryClass = "astro-ph.CO",
    doi = "10.1038/s42254-019-0137-0",
    journal = "Nature Rev. Phys.",
    volume = "2",
    number = "1",
    pages = "10--12",
    year = "2019"
}

@article{DiValentino:2020vnx,
    author = "Di Valentino, Eleonora",
    title = "{A combined analysis of the $H_0$ late time direct measurements and the impact on the Dark Energy sector}",
    eprint = "2011.00246",
    archivePrefix = "arXiv",
    primaryClass = "astro-ph.CO",
    reportNumber = "IPPP/20/72",
    doi = "10.1093/mnras/stab187",
    journal = "Mon. Not. Roy. Astron. Soc.",
    volume = "502",
    number = "2",
    pages = "2065--2073",
    year = "2021"
}

@article{Dominguez:2019jqc,
    author = "Dom{\'\i}nguez, Alberto and Wojtak, Radoslaw and Finke, Justin and Ajello, Marco and Helgason, Kari and Prada, Francisco and Desai, Abhishek and Paliya, Vaidehi and Marcotulli, Lea and Hartmann, Dieter",
    title = "{A new measurement of the Hubble constant and matter content of the Universe using extragalactic background light $\gamma$-ray attenuation}",
    eprint = "1903.12097",
    archivePrefix = "arXiv",
    primaryClass = "astro-ph.CO",
    doi = "10.3847/1538-4357/ab4a0e",
    journal = "Astrophys. J.",
    volume = "885",
    pages = "137",
    year = "2019"
}

@article{Boruah:2020fhl,
    author = "Boruah, Supranta S. and Hudson, Michael J. and Lavaux, Guilhem",
    title = "{Peculiar velocities in the local Universe: comparison of different models and the implications for H0 and dark matter}",
    eprint = "2010.01119",
    archivePrefix = "arXiv",
    primaryClass = "astro-ph.CO",
    doi = "10.1093/mnras/stab2320",
    journal = "Mon. Not. Roy. Astron. Soc.",
    volume = "507",
    number = "2",
    pages = "2697--2713",
    year = "2021"
}

@article{Mortsell:2021nzg,
    author = "Mortsell, Edvard and Goobar, Ariel and Johansson, Joel and Dhawan, Suhail",
    title = "{Sensitivity of the Hubble Constant Determination to Cepheid Calibration}",
    eprint = "2105.11461",
    archivePrefix = "arXiv",
    primaryClass = "astro-ph.CO",
    doi = "10.3847/1538-4357/ac756e",
    journal = "Astrophys. J.",
    volume = "933",
    number = "2",
    pages = "212",
    year = "2022"
}

@article{Mortsell:2021tcx,
    author = "Mortsell, Edvard and Goobar, Ariel and Johansson, Joel and Dhawan, Suhail",
    title = "{The Hubble Tension Revisited: Additional Local Distance Ladder Uncertainties}",
    eprint = "2106.09400",
    archivePrefix = "arXiv",
    primaryClass = "astro-ph.CO",
    doi = "10.3847/1538-4357/ac7c19",
    journal = "Astrophys. J.",
    volume = "935",
    number = "1",
    pages = "58",
    year = "2022"
}

@article{Sharon:2023ioz,
    author = "Sharon, Amir and Kushnir, Doron and Yuan, Wenlong and Macri, Lucas and Riess, Adam",
    title = "{Reassessing the constraints from SH0ES extragalactic Cepheid amplitudes on systematic blending bias}",
    eprint = "2305.14435",
    archivePrefix = "arXiv",
    primaryClass = "astro-ph.CO",
    doi = "10.1093/mnras/stae451",
    journal = "Mon. Not. Roy. Astron. Soc.",
    volume = "528",
    number = "4",
    pages = "6861--6880",
    year = "2024"
}

@article{Riess:2023bfx,
    author = "Riess, Adam G. and Anand, Gagandeep S. and Yuan, Wenlong and Casertano, Stefano and Dolphin, Andrew and Macri, Lucas M. and Breuval, Louise and Scolnic, Dan and Perrin, Marshall and Anderson, Richard I.",
    title = "{Crowded No More: The Accuracy of the Hubble Constant Tested with High-resolution Observations of Cepheids by JWST}",
    eprint = "2307.15806",
    archivePrefix = "arXiv",
    primaryClass = "astro-ph.CO",
    doi = "10.3847/2041-8213/acf769",
    journal = "Astrophys. J. Lett.",
    volume = "956",
    number = "1",
    pages = "L18",
    year = "2023"
}

@article{Bhardwaj:2023mau,
    author = "Bhardwaj, Anupam and others",
    title = "{High-resolution Spectroscopic Metallicities of Milky Way Cepheid Standards and Their Impact on the Leavitt Law and the Hubble Constant}",
    eprint = "2309.03263",
    archivePrefix = "arXiv",
    primaryClass = "astro-ph.SR",
    doi = "10.3847/2041-8213/acf710",
    journal = "Astrophys. J. Lett.",
    volume = "955",
    number = "1",
    pages = "L13",
    year = "2023"
}

@article{Brout:2023wol,
    author = "Brout, Dillon and Riess, Adam",
    title = "{The Impact of Dust on Cepheid and Type Ia Supernova Distances}",
    eprint = "2311.08253",
    archivePrefix = "arXiv",
    primaryClass = "astro-ph.CO",
    month = "11",
    year = "2023"
}

@article{Dwomoh:2023bro,
    author = "Dwomoh, Arianna M. and Peterson, Erik R. and Scolnic, Daniel and Ashall, Chris and DerKacy, James M. and Do, Aaron and Johansson, Joel and Jones, David O. and Riess, Adam G. and Shappee, Benjamin J.",
    title = "{Evaluating the Consistency of Cosmological Distances Using Supernova Siblings in the Near-infrared}",
    eprint = "2311.06178",
    archivePrefix = "arXiv",
    primaryClass = "astro-ph.CO",
    doi = "10.3847/1538-4357/ad1ff5",
    journal = "Astrophys. J.",
    volume = "965",
    number = "1",
    pages = "90",
    year = "2024"
}

@article{Riess:2024ohe,
    author = "Riess, Adam G. and Anand, Gagandeep S. and Yuan, Wenlong and Casertano, Stefano and Dolphin, Andrew and Macri, Lucas M. and Breuval, Louise and Scolnic, Dan and Perrin, Marshall and Anderson, I. Richard",
    title = "{JWST Observations Reject Unrecognized Crowding of Cepheid Photometry as an Explanation for the Hubble Tension at 8{\ensuremath{\sigma}} Confidence}",
    eprint = "2401.04773",
    archivePrefix = "arXiv",
    primaryClass = "astro-ph.CO",
    doi = "10.3847/2041-8213/ad1ddd",
    journal = "Astrophys. J. Lett.",
    volume = "962",
    number = "1",
    pages = "L17",
    year = "2024"
}

@article{DESI:2024mwx,
    author = "Adame, A. G. and others",
    collaboration = "DESI",
    title = "{DESI 2024 VI: cosmological constraints from the measurements of baryon acoustic oscillations}",
    eprint = "2404.03002",
    archivePrefix = "arXiv",
    primaryClass = "astro-ph.CO",
    reportNumber = "FERMILAB-PUB-24-0154-PPD",
    doi = "10.1088/1475-7516/2025/02/021",
    journal = "JCAP",
    volume = "02",
    pages = "021",
    year = "2025"
}

@article{DESI:2025zgx,
    author = "Abdul Karim, M. and others",
    collaboration = "DESI",
    title = "{DESI DR2 results. II. Measurements of baryon acoustic oscillations and cosmological constraints}",
    eprint = "2503.14738",
    archivePrefix = "arXiv",
    primaryClass = "astro-ph.CO",
    reportNumber = "FERMILAB-PUB-25-0169-PPD",
    doi = "10.1103/tr6y-kpc6",
    journal = "Phys. Rev. D",
    volume = "112",
    number = "8",
    pages = "083515",
    year = "2025"
}

@article{DES:2024jxu,
    author = "Abbott, T. M. C. and others",
    collaboration = "DES",
    title = "{The Dark Energy Survey: Cosmology Results with {\ensuremath{\sim}}1500 New High-redshift Type Ia Supernovae Using the Full 5 yr Data Set}",
    eprint = "2401.02929",
    archivePrefix = "arXiv",
    primaryClass = "astro-ph.CO",
    reportNumber = "FERMILAB-PUB-23-0821-PPD, DES-2023-805",
    doi = "10.3847/2041-8213/ad6f9f",
    journal = "Astrophys. J. Lett.",
    volume = "973",
    number = "1",
    pages = "L14",
    year = "2024"
}

@article{Chevallier:2000qy,
    author = "Chevallier, Michel and Polarski, David",
    title = "{Accelerating universes with scaling dark matter}",
    eprint = "gr-qc/0009008",
    archivePrefix = "arXiv",
    doi = "10.1142/S0218271801000822",
    journal = "Int. J. Mod. Phys. D",
    volume = "10",
    pages = "213--224",
    year = "2001"
}

@article{Linder:2002et,
    author = "Linder, Eric V.",
    title = "{Exploring the expansion history of the universe}",
    eprint = "astro-ph/0208512",
    archivePrefix = "arXiv",
    doi = "10.1103/PhysRevLett.90.091301",
    journal = "Phys. Rev. Lett.",
    volume = "90",
    pages = "091301",
    year = "2003"
}

@article{Giare:2025pzu,
    author = "Giar{\`e}, William and Mahassen, Tariq and Di Valentino, Eleonora and Pan, Supriya",
    title = "{An overview of what current data can (and cannot yet) say about evolving dark energy}",
    eprint = "2502.10264",
    archivePrefix = "arXiv",
    primaryClass = "astro-ph.CO",
    doi = "10.1016/j.dark.2025.101906",
    journal = "Phys. Dark Univ.",
    volume = "48",
    pages = "101906",
    year = "2025"
}

@article{Giare:2024ocw,
    author = "Giar{\`e}, William",
    title = "{Dynamical dark energy beyond Planck? Constraints from multiple CMB probes, DESI BAO, and type-Ia supernovae}",
    eprint = "2409.17074",
    archivePrefix = "arXiv",
    primaryClass = "astro-ph.CO",
    doi = "10.1103/ss37-cxhn",
    journal = "Phys. Rev. D",
    volume = "112",
    number = "2",
    pages = "023508",
    year = "2025"
}

@article{Cortes:2024lgw,
    author = "Cort{\^e}s, Marina and Liddle, Andrew R.",
    title = "{Interpreting DESI's evidence for evolving dark energy}",
    eprint = "2404.08056",
    archivePrefix = "arXiv",
    primaryClass = "astro-ph.CO",
    doi = "10.1088/1475-7516/2024/12/007",
    journal = "JCAP",
    volume = "12",
    pages = "007",
    year = "2024"
}

@article{Shlivko:2024llw,
    author = "Shlivko, David and Steinhardt, Paul J.",
    title = "{Assessing observational constraints on dark energy}",
    eprint = "2405.03933",
    archivePrefix = "arXiv",
    primaryClass = "astro-ph.CO",
    doi = "10.1016/j.physletb.2024.138826",
    journal = "Phys. Lett. B",
    volume = "855",
    pages = "138826",
    year = "2024"
}

@article{Luongo:2024fww,
    author = "Luongo, Orlando and Muccino, Marco",
    title = "{Model-independent cosmographic constraints from DESI 2024}",
    eprint = "2404.07070",
    archivePrefix = "arXiv",
    primaryClass = "astro-ph.CO",
    doi = "10.1051/0004-6361/202450512",
    journal = "Astron. Astrophys.",
    volume = "690",
    pages = "A40",
    year = "2024"
}

@article{Yin:2024hba,
    author = "Yin, Wen",
    title = "{Cosmic clues: DESI, dark energy, and the cosmological constant problem}",
    eprint = "2404.06444",
    archivePrefix = "arXiv",
    primaryClass = "hep-ph",
    doi = "10.1007/JHEP05(2024)327",
    journal = "JHEP",
    volume = "05",
    pages = "327",
    year = "2024"
}

@article{Gialamas:2024lyw,
    author = {Gialamas, Ioannis D. and H{\"u}tsi, Gert and Kannike, Kristjan and Racioppi, Antonio and Raidal, Martti and Vasar, Martin and Veerm{\"a}e, Hardi},
    title = "{Interpreting DESI 2024 BAO: Late-time dynamical dark energy or a local effect?}",
    eprint = "2406.07533",
    archivePrefix = "arXiv",
    primaryClass = "astro-ph.CO",
    doi = "10.1103/PhysRevD.111.043540",
    journal = "Phys. Rev. D",
    volume = "111",
    number = "4",
    pages = "043540",
    year = "2025"
}

@article{Dinda:2024kjf,
    author = "Dinda, Bikash R.",
    title = "{A new diagnostic for the null test of dynamical dark energy in light of DESI 2024 and other BAO data}",
    eprint = "2405.06618",
    archivePrefix = "arXiv",
    primaryClass = "astro-ph.CO",
    doi = "10.1088/1475-7516/2024/09/062",
    journal = "JCAP",
    volume = "09",
    pages = "062",
    year = "2024"
}

@article{Najafi:2024qzm,
    author = "Najafi, Mahdi and Pan, Supriya and Di Valentino, Eleonora and Firouzjaee, Javad T.",
    title = "{Dynamical dark energy confronted with multiple CMB missions}",
    eprint = "2407.14939",
    archivePrefix = "arXiv",
    primaryClass = "astro-ph.CO",
    doi = "10.1016/j.dark.2024.101539",
    journal = "Phys. Dark Univ.",
    volume = "45",
    pages = "101539",
    year = "2024"
}

@article{Wang:2024dka,
    author = "Wang, Hao and Piao, Yun-Song",
    title = "{Dark energy in light of DESI DR1 and Hubble tension}",
    eprint = "2404.18579",
    archivePrefix = "arXiv",
    primaryClass = "astro-ph.CO",
    doi = "10.1016/j.physletb.2026.140180",
    journal = "Phys. Lett. B",
    volume = "873",
    pages = "140180",
    year = "2026"
}

@article{Ye:2024ywg,
    author = "Ye, Gen and Martinelli, Matteo and Hu, Bin and Silvestri, Alessandra",
    title = "{Hints of Nonminimally Coupled Gravity in DESI 2024 Baryon Acoustic Oscillation Measurements}",
    eprint = "2407.15832",
    archivePrefix = "arXiv",
    primaryClass = "astro-ph.CO",
    doi = "10.1103/PhysRevLett.134.181002",
    journal = "Phys. Rev. Lett.",
    volume = "134",
    number = "18",
    pages = "181002",
    year = "2025"
}

@article{Tada:2024znt,
    author = "Tada, Yuichiro and Terada, Takahiro",
    title = "{Quintessential interpretation of the evolving dark energy in light of DESI observations}",
    eprint = "2404.05722",
    archivePrefix = "arXiv",
    primaryClass = "astro-ph.CO",
    doi = "10.1103/PhysRevD.109.L121305",
    journal = "Phys. Rev. D",
    volume = "109",
    number = "12",
    pages = "L121305",
    year = "2024"
}

@article{Carloni:2024zpl,
    author = "Carloni, Youri and Luongo, Orlando and Muccino, Marco",
    title = "{Does dark energy really revive using DESI 2024 data?}",
    eprint = "2404.12068",
    archivePrefix = "arXiv",
    primaryClass = "astro-ph.CO",
    doi = "10.1103/PhysRevD.111.023512",
    journal = "Phys. Rev. D",
    volume = "111",
    number = "2",
    pages = "023512",
    year = "2025"
}

@article{Chan-GyungPark:2024mlx,
    author = "Park, Chan-Gyung and de Cruz P{\'e}rez, Javier and Ratra, Bharat",
    title = "{Using non-DESI data to confirm and strengthen the DESI 2024 spatially flat w0waCDM cosmological parametrization result}",
    eprint = "2405.00502",
    archivePrefix = "arXiv",
    primaryClass = "astro-ph.CO",
    doi = "10.1103/PhysRevD.110.123533",
    journal = "Phys. Rev. D",
    volume = "110",
    number = "12",
    pages = "123533",
    year = "2024"
}

@article{DESI:2024kob,
    author = "Lodha, K. and others",
    collaboration = "DESI",
    title = "{DESI 2024: Constraints on physics-focused aspects of dark energy using DESI DR1 BAO data}",
    eprint = "2405.13588",
    archivePrefix = "arXiv",
    primaryClass = "astro-ph.CO",
    reportNumber = "FERMILAB-PUB-24-0756-PPD",
    doi = "10.1103/PhysRevD.111.023532",
    journal = "Phys. Rev. D",
    volume = "111",
    number = "2",
    pages = "023532",
    year = "2025"
}

@article{Bhattacharya:2024hep,
    author = "Bhattacharya, Sukannya and Borghetto, Giulia and Malhotra, Ameek and Parameswaran, Susha and Tasinato, Gianmassimo and Zavala, Ivonne",
    title = "{Cosmological constraints on curved quintessence}",
    eprint = "2405.17396",
    archivePrefix = "arXiv",
    primaryClass = "astro-ph.CO",
    doi = "10.1088/1475-7516/2024/09/073",
    journal = "JCAP",
    volume = "09",
    pages = "073",
    year = "2024"
}

@article{Ramadan:2024kmn,
    author = "Ramadan, Omar F. and Sakstein, Jeremy and Rubin, David",
    title = "{DESI constraints on exponential quintessence}",
    eprint = "2405.18747",
    archivePrefix = "arXiv",
    primaryClass = "astro-ph.CO",
    doi = "10.1103/PhysRevD.110.L041303",
    journal = "Phys. Rev. D",
    volume = "110",
    number = "4",
    pages = "L041303",
    year = "2024"
}

@article{Notari:2024rti,
    author = "Notari, Alessio and Redi, Michele and Tesi, Andrea",
    title = "{Consistent theories for the DESI dark energy fit}",
    eprint = "2406.08459",
    archivePrefix = "arXiv",
    primaryClass = "astro-ph.CO",
    doi = "10.1088/1475-7516/2024/11/025",
    journal = "JCAP",
    volume = "11",
    pages = "025",
    year = "2024"
}

@article{Orchard:2024bve,
    author = "Orchard, Lili and C{\'a}rdenas, V{\'\i}ctor H.",
    title = "{Probing dark energy evolution post-DESI 2024}",
    eprint = "2407.05579",
    archivePrefix = "arXiv",
    primaryClass = "astro-ph.CO",
    doi = "10.1016/j.dark.2024.101678",
    journal = "Phys. Dark Univ.",
    volume = "46",
    pages = "101678",
    year = "2024"
}

@article{Hernandez-Almada:2024ost,
    author = "Hern{\'a}ndez-Almada, A. and Mendoza-Mart{\'\i}nez, M. L. and Garc{\'\i}a-Aspeitia, Miguel A. and Motta, V.",
    title = "{Phenomenological emergent dark energy in the light of DESI Data Release 1}",
    eprint = "2407.09430",
    archivePrefix = "arXiv",
    primaryClass = "astro-ph.CO",
    doi = "10.1016/j.dark.2024.101668",
    journal = "Phys. Dark Univ.",
    volume = "46",
    pages = "101668",
    year = "2024"
}

@article{Pourojaghi:2024tmw,
    author = "Malekjani, Mohammad and Davari, Zahra and Pourojaghi, Saeed",
    collaboration = "DESI",
    title = "{Cosmological constraints on dark energy parametrizations after DESI 2024: Persistent deviation from standard {\ensuremath{\Lambda}}CDM cosmology}",
    eprint = "2407.09767",
    archivePrefix = "arXiv",
    primaryClass = "astro-ph.CO",
    doi = "10.1103/PhysRevD.111.083547",
    journal = "Phys. Rev. D",
    volume = "111",
    number = "8",
    pages = "083547",
    year = "2025"
}

@article{Giare:2024gpk,
    author = "Giar{\`e}, William and Najafi, Mahdi and Pan, Supriya and Di Valentino, Eleonora and Firouzjaee, Javad T.",
    title = "{Robust preference for Dynamical Dark Energy in DESI BAO and SN measurements}",
    eprint = "2407.16689",
    archivePrefix = "arXiv",
    primaryClass = "astro-ph.CO",
    doi = "10.1088/1475-7516/2024/10/035",
    journal = "JCAP",
    volume = "10",
    pages = "035",
    year = "2024"
}

@article{Reboucas:2024smm,
    author = "Rebou{\c{c}}as, Jo{\~a}o and de Souza, Diogo H. F. and Zhong, Kunhao and Miranda, Vivian and Rosenfeld, Rogerio",
    title = "{Investigating late-time dark energy and massive neutrinos in light of DESI Y1 BAO}",
    eprint = "2408.14628",
    archivePrefix = "arXiv",
    primaryClass = "astro-ph.CO",
    doi = "10.1088/1475-7516/2025/02/024",
    journal = "JCAP",
    volume = "02",
    pages = "024",
    year = "2025"
}

@article{Chan-GyungPark:2024brx,
    author = "Park, Chan-Gyung and de Cruz P{\'e}rez, Javier and Ratra, Bharat",
    title = "{Is the w0waCDM cosmological parameterization evidence for dark energy dynamics partially caused by the excess smoothing of Planck CMB anisotropy data?}",
    eprint = "2410.13627",
    archivePrefix = "arXiv",
    primaryClass = "astro-ph.CO",
    doi = "10.1142/S0218271825500580",
    journal = "Int. J. Mod. Phys. D",
    volume = "34",
    number = "13",
    pages = "2550058",
    year = "2025"
}

@article{Menci:2024hop,
    author = "Menci, Nicola and Sen, Anjan Ananda and Castellano, Marco",
    title = "{The Excess of JWST Bright Galaxies: A Possible Origin in the Ground State of Dynamical Dark Energy in the Light of DESI 2024 Data}",
    eprint = "2410.22940",
    archivePrefix = "arXiv",
    primaryClass = "astro-ph.CO",
    doi = "10.3847/1538-4357/ad8d5b",
    journal = "Astrophys. J.",
    volume = "976",
    number = "2",
    pages = "227",
    year = "2024"
}

@article{Li:2024qus,
    author = "Li, Tian-Nuo and Li, Yun-He and Du, Guo-Hong and Wu, Peng-Ju and Feng, Lu and Zhang, Jing-Fei and Zhang, Xin",
    title = "{Revisiting holographic dark energy after DESI 2024}",
    eprint = "2411.08639",
    archivePrefix = "arXiv",
    primaryClass = "astro-ph.CO",
    doi = "10.1140/epjc/s10052-025-14279-7",
    journal = "Eur. Phys. J. C",
    volume = "85",
    number = "6",
    pages = "608",
    year = "2025"
}

@article{Li:2024hrv,
    author = "Li, Jun-Xian and Wang, Shuang",
    title = "{A comprehensive numerical study on four categories of holographic dark energy models}",
    eprint = "2412.09064",
    archivePrefix = "arXiv",
    primaryClass = "astro-ph.CO",
    doi = "10.1088/1475-7516/2025/07/047",
    journal = "JCAP",
    volume = "07",
    pages = "047",
    year = "2025"
}

@article{Notari:2024zmi,
    author = "Notari, Alessio and Redi, Michele and Tesi, Andrea",
    title = "{BAO vs. SN evidence for evolving dark energy}",
    eprint = "2411.11685",
    archivePrefix = "arXiv",
    primaryClass = "astro-ph.CO",
    doi = "10.1088/1475-7516/2025/04/048",
    journal = "JCAP",
    volume = "04",
    pages = "048",
    year = "2025"
}

@article{Gao:2024ily,
    author = "Gao, Qing and Peng, Zhiqian and Gao, Shengqing and Gong, Yungui",
    title = "{On the Evidence of Dynamical Dark Energy}",
    eprint = "2411.16046",
    archivePrefix = "arXiv",
    primaryClass = "astro-ph.CO",
    doi = "10.3390/universe11010010",
    journal = "Universe",
    volume = "11",
    number = "1",
    pages = "10",
    year = "2025"
}

@article{Fikri:2024klc,
    author = "Fikri, Ramy and Elkhateeb, Esraa and Lashin, E. I. and El Hanafy, Waleed",
    title = "{A preference for dynamical phantom dark energy using one-parameter model with Planck, DESI DR1 BAO and SN data}",
    eprint = "2411.19362",
    archivePrefix = "arXiv",
    primaryClass = "astro-ph.CO",
    doi = "10.1016/j.aop.2025.170190",
    journal = "Annals Phys.",
    volume = "481",
    pages = "170190",
    year = "2025"
}

@article{Jiang:2024xnu,
    author = "Jiang, Jun-Qian and Pedrotti, Davide and da Costa, Simony Santos and Vagnozzi, Sunny",
    title = "{Nonparametric late-time expansion history reconstruction and implications for the Hubble tension in light of recent DESI and type Ia supernovae data}",
    eprint = "2408.02365",
    archivePrefix = "arXiv",
    primaryClass = "astro-ph.CO",
    doi = "10.1103/PhysRevD.110.123519",
    journal = "Phys. Rev. D",
    volume = "110",
    number = "12",
    pages = "123519",
    year = "2024"
}

@article{Zheng:2024qzi,
    author = "Zheng, Jie and Qiang, Da-Chun and You, Zhi-Qiang",
    title = "{Cosmological constraints on dark energy models using DESI BAO 2024}",
    eprint = "2412.04830",
    archivePrefix = "arXiv",
    primaryClass = "astro-ph.CO",
    doi = "10.1088/1475-7516/2025/08/056",
    journal = "JCAP",
    volume = "08",
    pages = "056",
    year = "2025"
}

@article{Gomez-Valent:2024ejh,
    author = "G{\'o}mez-Valent, Adria and Sol{\`a} Peracaula, Joan",
    title = "{Composite dark energy and the cosmological tensions}",
    eprint = "2412.15124",
    archivePrefix = "arXiv",
    primaryClass = "astro-ph.CO",
    doi = "10.1016/j.physletb.2025.139391",
    journal = "Phys. Lett. B",
    volume = "864",
    pages = "139391",
    year = "2025"
}

@article{RoyChoudhury:2024wri,
    author = "Roy Choudhury, Shouvik and Okumura, Teppei",
    title = "{Updated Cosmological Constraints in Extended Parameter Space with Planck PR4, DESI Baryon Acoustic Oscillations, and Supernovae: Dynamical Dark Energy, Neutrino Masses, Lensing Anomaly, and the Hubble Tension}",
    eprint = "2409.13022",
    archivePrefix = "arXiv",
    primaryClass = "astro-ph.CO",
    doi = "10.3847/2041-8213/ad8c26",
    journal = "Astrophys. J. Lett.",
    volume = "976",
    number = "1",
    pages = "L11",
    year = "2024"
}

@article{Lewis:2024cqj,
    author = "Lewis, Antony and Chamberlain, Ewan",
    title = "{Understanding acoustic scale observations: the one-sided fight against {\ensuremath{\Lambda}}}",
    eprint = "2412.13894",
    archivePrefix = "arXiv",
    primaryClass = "astro-ph.CO",
    doi = "10.1088/1475-7516/2025/05/065",
    journal = "JCAP",
    volume = "05",
    pages = "065",
    year = "2025"
}

@article{Wolf:2025jlc,
    author = "Wolf, William J. and Garc{\'\i}a-Garc{\'\i}a, Carlos and Ferreira, Pedro G.",
    title = "{Robustness of dark energy phenomenology across different parameterizations}",
    eprint = "2502.04929",
    archivePrefix = "arXiv",
    primaryClass = "astro-ph.CO",
    doi = "10.1088/1475-7516/2025/05/034",
    journal = "JCAP",
    volume = "05",
    pages = "034",
    year = "2025"
}

@article{Shajib:2025tpd,
    author = "Shajib, Anowar J. and Frieman, Joshua A.",
    title = "{Scalar-field dark energy models: Current and forecast constraints}",
    eprint = "2502.06929",
    archivePrefix = "arXiv",
    primaryClass = "astro-ph.CO",
    doi = "10.1103/kjpb-r698",
    journal = "Phys. Rev. D",
    volume = "112",
    number = "6",
    pages = "063508",
    year = "2025"
}

@article{Chaussidon:2025npr,
    author = "Chaussidon, E. and others",
    title = "{Early time solution as an alternative to the late time evolving dark energy with DESI DR2 BAO}",
    eprint = "2503.24343",
    archivePrefix = "arXiv",
    primaryClass = "astro-ph.CO",
    reportNumber = "FERMILAB-PUB-25-0241-PPD",
    doi = "10.1103/xtql-wh3h",
    journal = "Phys. Rev. D",
    volume = "112",
    number = "6",
    pages = "063548",
    year = "2025"
}

@article{Kessler:2025kju,
    author = "Kessler, Daniel A. and Escamilla, Luis A. and Pan, Supriya and Di Valentino, Eleonora",
    title = "{One-parameter dynamical dark energy: Hints for oscillations}",
    eprint = "2504.00776",
    archivePrefix = "arXiv",
    primaryClass = "astro-ph.CO",
    doi = "10.1103/1x6w-4tnd",
    journal = "Phys. Rev. D",
    volume = "114",
    number = "6",
    pages = "063501",
    year = "2026"
}

@article{Pang:2025lvh,
    author = "Pang, Ye-Huang and Zhang, Xue and Huang, Qing-Guo",
    title = "{The impact of the Hubble tension on the evidence for dynamical dark energy}",
    eprint = "2503.21600",
    archivePrefix = "arXiv",
    primaryClass = "astro-ph.CO",
    doi = "10.1007/s11433-025-2713-8",
    journal = "Sci. China Phys. Mech. Astron.",
    volume = "68",
    number = "8",
    pages = "280410",
    year = "2025"
}

@article{RoyChoudhury:2025dhe,
    author = "Roy Choudhury, Shouvik",
    title = "{Cosmology in Extended Parameter Space with DESI Data Release 2 Baryon Acoustic Oscillations: A 2{\ensuremath{\sigma}}+ Detection of Nonzero Neutrino Masses with an Update on Dynamical Dark Energy and Lensing Anomaly}",
    eprint = "2504.15340",
    archivePrefix = "arXiv",
    primaryClass = "astro-ph.CO",
    doi = "10.3847/2041-8213/ade1cc",
    journal = "Astrophys. J. Lett.",
    volume = "986",
    number = "2",
    pages = "L31",
    year = "2025",
    note = "[Erratum: Astrophys.J.Lett. 1001, L25 (2026), Erratum: Astrophys.J. 1001, L25 (2026)]"
}

@article{Scherer:2025esj,
    author = "Scherer, Mateus and Sabogal, Miguel A. and Nunes, Rafael C. and De Felice, Antonio",
    title = "{Challenging the {\ensuremath{\Lambda}}CDM model: 5{\ensuremath{\sigma}} evidence for a dynamical dark energy late-time transition}",
    eprint = "2504.20664",
    archivePrefix = "arXiv",
    primaryClass = "astro-ph.CO",
    doi = "10.1103/n86r-sjgm",
    journal = "Phys. Rev. D",
    volume = "112",
    number = "4",
    pages = "043513",
    year = "2025"
}

@article{Specogna:2025guo,
    author = "Specogna, Enrico and Adil, Shahnawaz A. and Ozulker, Emre and Di Valentino, Eleonora and Nunes, Rafael C. and Akarsu, Ozgur and Sen, Anjan A.",
    title = "{Updated constraints on omnipotent dark energy: A comprehensive analysis with CMB and BAO data}",
    eprint = "2504.17859",
    archivePrefix = "arXiv",
    primaryClass = "gr-qc",
    doi = "10.1103/b7ht-lx26",
    journal = "Phys. Rev. D",
    volume = "113",
    number = "10",
    pages = "103549",
    year = "2026"
}

@article{Cheng:2025lod,
    author = "Cheng, Hanyu and Di Valentino, Eleonora and Escamilla, Luis A. and Sen, Anjan A. and Visinelli, Luca",
    title = "{Pressure parametrization of dark energy: first and second-order constraints with latest cosmological data}",
    eprint = "2505.02932",
    archivePrefix = "arXiv",
    primaryClass = "astro-ph.CO",
    reportNumber = "CA21106; CA21136",
    doi = "10.1088/1475-7516/2025/09/031",
    journal = "JCAP",
    volume = "09",
    pages = "031",
    year = "2025"
}

@article{Nesseris:2025lke,
    author = "Nesseris, Savvas and Akrami, Yashar and Starkman, Glenn D.",
    title = "{To CPL, or not to CPL? What we have not learned about the dark energy equation of state}",
    eprint = "2503.22529",
    archivePrefix = "arXiv",
    primaryClass = "astro-ph.CO",
    reportNumber = "IFT-UAM/CSIC-25-29",
    doi = "10.1103/q2gb-n61g",
    journal = "Phys. Rev. D",
    volume = "114",
    number = "4",
    pages = "L041305",
    year = "2026"
}

@article{Akrami:2025zlb,
    author = "Akrami, Yashar and Alestas, George and Nesseris, Savvas",
    title = "{Has DESI detected exponential quintessence?}",
    eprint = "2504.04226",
    archivePrefix = "arXiv",
    primaryClass = "astro-ph.CO",
    reportNumber = "IFT-UAM/CSIC-25-36",
    month = "4",
    year = "2025"
}

@article{Cheng:2025hug,
    author = "Cheng, Hanyu and Di Valentino, Eleonora and Visinelli, Luca",
    title = "{Cosmic strings as dynamical dark energy: Novel constraints}",
    eprint = "2505.22066",
    archivePrefix = "arXiv",
    primaryClass = "astro-ph.CO",
    doi = "10.1016/j.jheap.2026.100610",
    journal = "JHEAp",
    volume = "53",
    pages = "100610",
    year = "2026"
}

@article{Gonzalez-Fuentes:2025lei,
    author = "Gonz{\'a}lez-Fuentes, Alex and G{\'o}mez-Valent, Adri{\`a}",
    title = "{Reconstruction of dark energy and late-time cosmic expansion using the Weighted Function Regression method}",
    eprint = "2506.11758",
    archivePrefix = "arXiv",
    primaryClass = "astro-ph.CO",
    doi = "10.1088/1475-7516/2025/12/049",
    journal = "JCAP",
    volume = "12",
    pages = "049",
    year = "2025"
}

@article{Ye:2024zpk,
    author = "Ye, Gen",
    title = "{Modified gravity bridges the cosmological tensions}",
    eprint = "2411.11743",
    archivePrefix = "arXiv",
    primaryClass = "astro-ph.CO",
    doi = "10.1103/2zc8-m8s8",
    journal = "Phys. Rev. D",
    volume = "114",
    number = "6",
    pages = "L061301",
    year = "2026"
}

@article{Sabogal:2025jbo,
    author = "Sabogal, Miguel A. and Nunes, Rafael C.",
    title = "{Robust evidence for dynamical dark energy from DESI galaxy-CMB lensing cross-correlation and geometric probes}",
    eprint = "2505.24465",
    archivePrefix = "arXiv",
    primaryClass = "astro-ph.CO",
    doi = "10.1088/1475-7516/2025/09/084",
    journal = "JCAP",
    volume = "09",
    pages = "084",
    year = "2025"
}

@article{Gomez-Valent:2025mfl,
    author = "G{\'o}mez-Valent, Adri{\`a} and Gonz{\'a}lez-Fuentes, Alex",
    title = "{Effective phantom divide crossing with standard and negative quintessence}",
    eprint = "2508.00621",
    archivePrefix = "arXiv",
    primaryClass = "astro-ph.CO",
    doi = "10.1016/j.physletb.2025.140096",
    journal = "Phys. Lett. B",
    volume = "872",
    pages = "140096",
    year = "2026"
}

@article{Lu:2025gki,
    author = "Lu, Zhiyu and Simon, Th{\'e}o and Zhang, Pierre",
    title = "{Preference for evolving dark energy in light of the galaxy bispectrum}",
    eprint = "2503.04602",
    archivePrefix = "arXiv",
    primaryClass = "astro-ph.CO",
    month = "3",
    year = "2025"
}

@article{Cortes:2025joz,
    author = "Cort{\^e}s, Marina and Liddle, Andrew R.",
    title = "{On DESI's DR2 exclusion of $Λ$CDM}",
    eprint = "2504.15336",
    archivePrefix = "arXiv",
    primaryClass = "astro-ph.CO",
    doi = "10.1093/mnrasl/slaf108",
    journal = "Mon. Not. Roy. Astron. Soc.",
    volume = "544",
    pages = "L121--L125",
    year = "2025"
}

@article{Wang:2025bkk,
    author = "Wang, Deng and Mota, David",
    title = "{Did DESI DR2 truly reveal dynamical dark energy?}",
    eprint = "2504.15222",
    archivePrefix = "arXiv",
    primaryClass = "astro-ph.CO",
    doi = "10.1140/epjc/s10052-025-15076-y",
    journal = "Eur. Phys. J. C",
    volume = "85",
    number = "11",
    pages = "1356",
    year = "2025"
}

@article{Keeley:2025rlg,
    author = "Keeley, Ryan E. and Shafieloo, Arman and Matthewson, William L.",
    title = "{Could We Be Fooled about Phantom Crossing?}",
    eprint = "2506.15091",
    archivePrefix = "arXiv",
    primaryClass = "astro-ph.CO",
    month = "6",
    year = "2025"
}
\end{document}